\documentclass[epsf,amssymb,aps,prd,nofootinbib,floatfix,superscriptaddress,secnumarabic]{revtex4}
\usepackage{bm}
\usepackage{amsmath}                                                                                          
\usepackage{graphicx}%

\usepackage{epsf}
\usepackage{epsfig}

 \usepackage{amsmath}
\usepackage{amsfonts,amssymb}

\usepackage{colordvi}
\usepackage{color}
\usepackage{lscape}
\usepackage{rotfloat}
\usepackage{rotating}

\usepackage{dsfont}
\usepackage{slashed}
\usepackage{booktabs}

\newcommand{\be}{\begin{equation}}
\newcommand{\ee}{\end{equation}}
\newcommand{\bea}{\begin{eqnarray}}
\newcommand{\eea}{\end{eqnarray}}

\newcommand{\MSbar}{{\overline{\rm MS}}}
\newcommand{\pa}{\partial}
\newcommand{\gtilde}{\frac{g^2}{16 \, \pi^2}\; }
\newcommand{\qslash}{{\not{\hspace{-0.05cm}q}}}
\newcommand{\thb}{\bar{\theta}}
\newcommand{\la}{\lambda}
\newcommand{\si}{\sigma}
\newcommand{\al}{\alpha}

\newcommand{\bpsi}{\overline{\psi}}

\def\lsim{\mathrel{\rlap{\lower4pt\hbox{\hskip1pt$\sim$}}
    \raise1pt\hbox{$<$}}}                
\def\slashed{{/}\mskip-10.0mu}
\def\qcircslash{\slashed {q\mskip -5mu ^{^\circ}}}

\begin{document}
\vspace*{1.75cm}

\title{Supersymmetric QCD on the Lattice: An Exploratory Study}

\author{M.~Costa$^{\ast 1}$ and H.~Panagopoulos\footnote{Electronic address: kosta.marios@ucy.ac.cy, haris@ucy.ac.cy}}
\affiliation{Department of Physics, University of Cyprus, Nicosia, CY-1678, Cyprus}

\begin{abstract}
We perform a pilot study of the perturbative renormalization of a Supersymmetric gauge theory with matter fields on the lattice. As a specific example, we consider Supersymmetric ${\cal N}{=}1$ QCD (SQCD). We study the self-energies of all particles which appear in this theory, as well as the renormalization of the coupling constant. To this end we compute, perturbatively to one-loop, the relevant two-point and three-point Green's functions using both dimensional and lattice regularizations. Our lattice formulation involves the Wilson discretization for the gluino and quark fields; for gluons we employ the Wilson gauge action; for scalar fields (squarks) we use na\"ive discretization. The gauge group that we consider is $SU(N_c)$, while the number of colors, $N_c$, the number of flavors, $N_f$, and the gauge parameter, $\alpha$, are left unspecified.

We obtain analytic expressions for the renormalization factors of the coupling constant ($Z_g$) and of the quark ($Z_\psi$), gluon ($Z_u$), gluino ($Z_\lambda$), squark ($Z_{A_\pm}$), and ghost ($Z_c$) fields on the lattice. We also compute the critical values of the gluino, quark and squark masses. Finally, we address the mixing which occurs among squark degrees of freedom beyond tree level: we calculate the corresponding mixing matrix which is necessary in order to disentangle the components of the squark field via an additional finite renormalization.
\end{abstract}

\maketitle

\section{Introduction}
\bigskip

In recent years the prospects of extracting nonperturbative information for Supersymmetric Theories through lattice simulations are being addressed extensively, from a number of viewpoints \cite{Catterall:2011, Suzuki&Tani:2005, Giet&Poppitz, Feo:2003, Kaplan:2009, Creutz:2001, Feo:2013, Catterall:2014}. There are a number of important physical questions regarding Supersymmetry (SUSY) to be ultimately investigated on the lattice, such as the nature of SUSY breaking, and the phase diagram of SUSY models. Such questions have become increasingly relevant in recent years, in the context of studies of Beyond-the-Standard-Model (BSM) Physics. Many notorious problems arise when formulating SUSY models on the lattice, such as the emergence of a plethora of counterterms in the action and the need for fine-tuning of masses and coupling constants \cite{Giet&Poppitz}. The present paper investigates these problems using, as a representative nontrivial model, supersymmetric $SU(N_c)$ Quantum Chromodynamics (SQCD), with ${\cal N}{=}1$ supersymmetric generators and $N_f$ flavors of matter fields.

Regularizing a Field Theory on the lattice entails breaking several symmetries, including Lorentz/rotational symmetry, chiral symmetry and, inevitably, Supersymmetry. Depending on  the type of discretization, a small subset of the original SUSY generators may be left intact in some models; the study of such cases is very interesting on its own merits \cite{Kaplan:2009}. However, in most models of interest, SUSY is thoroughly broken upon discretization; thus one must carefully assess the possibility of restoring the symmetry in the continuum limit. In the 
absence of anomalies, symmetry restoration amounts to introducing appropriate counterterms to the regularized Lagrangian, and thus it is a feasible procedure in principle; the correctness of such a procedure was established, for Supersymmetric Yang-Mills theories, in the seminal paper of Curci and Veneziano \cite{Curci&Venz}. Nevertheless, a satisfactory calibration of counterterms appears to be a formidable problem; not only can their number be exceedingly large \cite{Giedt}, but some of them can diverge with inverse powers of the lattice spacing, making their determination intractable by perturbation theory alone. At best, one must evoke both perturbative and non-perturbative methods \cite{josephREVIEW} in order to achieve a reliable renormalization of the theory. The present work represents a first step towards this goal.

In a supersymmetric version of QCD, both gluons and quark fields are promoted to superfields, containing fermionic and bosonic components. In this paper we concentrate on ${\cal N} = 1$ supersymmetry in the Wess-Zumino (WZ) gauge. In this gauge, the SQCD Lagrangian contains the following fields: the gluon together with the gluino and one real auxiliary scalar; in addition, for each quark flavor, a Dirac fermion, two squarks and two complex auxiliary scalars. The squark fields are complex scalar bosons, and the gluino field is a Majorana fermion which mediates interactions of the squark fields with their fermionic partners. In addition to these interactions, the gluon field is coupled with all superpartners and with the quark field. Since all these interactions must have the same coupling, the theory has a unique coupling constant renormalization. Furthermore, SUSY requires that the renormalized masses for quark and squark fields be the same.

The outline of this paper is as follows: Section~\ref{sec2} provides a brief theoretical background in which we introduce the supersymmetric trasformation of all fields which appear in the classical Lagrangian of SQCD. The main part of this study is Section~\ref{sec3} which contains a summary of the calculational procedure for the two-point (2-pt) Green's functions for each field and three point (3-pt) Green's function for the determination of the coupling constant. This Section is divided in two subsections: In Subsection~\ref{sec3.1}, we calculate the 2-pt and 3-pt Green's functions using dimensional regularization and in Subsection~\ref{sec3.2}, using the lattice formulation. Furthermore, in Section~\ref{sec3.2}, using the results that we found in Section~\ref{sec3.1}, we extract the renormalization of all fields and of the coupling constant, in the $\MSbar$ (modified minimal subtraction) renormalization scheme. We also discuss the finite mixing between the two squark fields, which appears on the lattice but also in dimensional regularization, depending on the $D$-dimensional prescription for $\gamma_5$, and calculate the corresponding mixing matrix. Finally, we conclude in Section~\ref{summary} with a discussion of our results and possible future extensions of our work.
 
\section{The continuum Action and Transformation Properties of the component fields}
\label{sec2}

Superfields are functions over superspace $\{x^\mu, \theta, \thb\}$, where $\theta$, $\thb$ are anticommuting variables and $x_\mu$ are the spacetime coordinates. The construction of the Lagrangian of SQCD involves chiral superfields and vector superfields. These fields are described in detail in, e.g., Refs.~\cite{Grisaru:1983, Wess&Bagger:1992,  Weinberg:2000,  MartinP}; in what follows, we list some of their properties, for the sake of a self-contained presentation. The physical components of a chiral superfield $\Phi$ are the matter fields: $A(x)$ which represents a complex scalar (squark), $\psi(x)$ which is a two-component spinor (quark - spin  $\frac{1}{2}$) and $F(x)$ which is an auxiliary complex scalar field. All components of the chiral superfield carry a color index in the fundamental representation of $SU(N_c)$. In superspace notation  the chiral superfield $\Phi$ in terms of the above component fields is:
\bea
    \Phi(x ;\theta ,\thb) &=& A(y) + \sqrt{2} \, \theta \psi(y)  +  \theta \theta \, F(y) \qquad\qquad (y^\mu \equiv x^\mu + i\theta \sigma^\mu \thb)
\nonumber\\ 
&=& A(x) + \sqrt{2} \, \theta \psi(x)  +  \theta \theta \, F(x) +
 i \theta \, \sigma^{\mu} \, \thb  \, \pa_{\mu} A(x) +  \frac{i}{\sqrt{2}} \theta \theta\, \thb \, \bar\sigma^{\mu} \, \pa_{\mu} \psi(x) + \frac{1}{4} \; \theta \theta \, \thb \thb \, \pa_{\mu} \pa^{\mu} A(x).
\eea 
The general form of a vector superfield $V(x,\theta, \thb)$ is:
\bea
V(x;\theta,\thb) &=&  C(x) +  i\theta \chi(x) -  i \thb \bar{\chi}(x)  +  \frac{i}{2} \theta \theta \, \big[M(x) \, + \, iN(x) \big] -  \frac{i}{2} \thb \thb \big[M(x) \, - \, iN(x) \big] \\ \nonumber
& - & \theta \, \sigma^{\mu} \, \thb \, u_{\mu}(x)  +  i \theta \theta \, \thb \left[ \bar{\la}(x) \, + \, \frac{i}{2} \bar{\si}^{\mu} \pa_{\mu} \chi(x) \right] -  i\thb \thb \, \theta \left[\la(x) \, + \, \frac{i}{2} \sigma^{\mu} \pa_{\mu} \bar{\chi}(x) \right] \\ \nonumber 
& + & \frac{1}{2} \, \theta \theta \, \thb \thb \, \left[D(x) \, + \, \frac{1}{2} \pa_{\mu} \pa^{\mu} C(x) \right],
\eea
where $C(x)$, $D(x)$, $M(x)$, $N(x)$ are real scalar fields, $\chi(x)$ and $\lambda(x)$ are two-component spinors, and $u_\mu(x)$ is a vector field; all components are in the adjoint representation of $SU(N_c)$: $V(x;\theta,\thb) = V(x;\theta,\thb)^\alpha T^\alpha $, where $T^\alpha$ are the generators of $SU(N_c)$ and $\alpha = 1, \ldots, N_c^2 - 1$. A supersymmetric gauge transformation may be applied on both the chiral and vector superfields, see Eq.~(\ref{SgaugeTran}), and we will require the Lagrangian to be invariant under this transformation.
We can choose a special gauge where the components $C , \chi , M ,N$ are zero. This defines the Wess-Zumino (WZ) gauge. A vector superfield in the WZ gauge reduces to the form:

\be    
V(x; \theta, \thb) = -\theta \, \si^{\mu} \, \thb \, u_{\mu}(x)  + i \theta \theta \, \thb \bar{\la}(x) - i \thb \thb \,\theta \la(x) + \ \frac{1}{2} \, \theta \theta \, \thb \thb \, D(x) \ ,
\ee
where $u_{\mu}^{\al}$ is the gluon field, $\la^{\al}$ is the gluino field and $D^{\al}$ is an auxiliary field.
\bigskip

In order to obtain a renormalizable theory, we need to construct a Lagrangian with products of superfields having dimensionality $\leq 4$; in addition, we require Lorentz invariance as well as invariance under supersymmetric gauge transformations:
\bea
\Phi'_+ &=& \rm{e}^{-i \Lambda} \Phi_+ \nonumber \\ 
\label{SgaugeTran}
\Phi'_- &=&  \Phi_- \rm{e}^{i \Lambda}\\ \nonumber
\rm{e}^{2g\,V'} &=& \rm{e}^{-i \Lambda^{\dagger}} \rm{e}^{2g\,V} \rm{e}^{i \Lambda},
\eea
where $\Lambda(x; \theta, \thb)$ is an arbitrary chiral superfield: $\Lambda(x; \theta, \thb) = \Lambda_0(y) + \sqrt{2} \theta \Lambda_1(y) + \theta \theta \Lambda_2(y)$. The special case 
in which $\Lambda_1 = \Lambda_2 = 0 \rightarrow \Lambda(x; \theta,\bar\theta) = \Lambda_0(y) = \Lambda_0(x) + i \theta \sigma^\mu \thb \partial_\mu \Lambda_0(x) + \frac{1}{4} \theta \theta \, \thb \thb\, \Box \Lambda_0(x) $, where  $\Lambda_0(x) = \Lambda_0^\dagger(x)$, amount to ordinary gauge transformations, which do not take us out of the WZ gauge: 
\begin{align}
A_+' &= G^{-1} A_+,& \psi_+'&= G^{-1} \psi_+,& F_+' &= G^{-1} F_+   \qquad\qquad(G(x) \equiv e^{i \Lambda_0(x)} )\nonumber \\
\label{SgaugeTranComponents}
A_-'&= A_- G,& \psi_-'&= \psi_- G,& F_-' &= F_- G \\\nonumber
u'_\mu &= G^{-1} u_\mu G + \frac{i}{g} (\partial_\mu G^{-1})G, &\lambda' &= G^{-1} \lambda G,& D' &=  G^{-1} D G.
\end{align}

A Lagrangian, which respects the trasformations of Eq.~(\ref{SgaugeTran}), in terms of superfields is:
\be
{\cal L} = \frac{1}{16 k g} {\rm{Tr}} \big( W^{\al} \, W_{\al}|_{\theta \theta} +\bar{W}_{\dot{\al}} \, \bar{W}^{\dot{\al}}|_{\thb \thb}\big) 
+\big( \Phi^{\dagger}_+ \, \rm{e}^{2g\,V} \, \Phi_+ +  \Phi_- \, \rm{e}^{-2g\,V} \, \Phi^{\dagger}_- \big)|_{\theta \theta \thb \thb}
+ m \big(\Phi_-\Phi_+|_{\theta \theta} +  \Phi^{\dagger}_+\Phi^{\dagger}_-|_{\thb \thb} \big),
\label{superfieldLag}
\ee

\noindent where ${\rm Tr} (T^{\alpha}T^{\beta}) = k \,\delta^{\alpha\,\beta}$, $W_{\al} =  -\frac{1}{4} \bar{{\cal D}}\bar{{\cal D}} \,e^{-2g\,V} \,  {\cal D}_{\al} \,e^{2g\,V} $ is the supersymmetric field strength, and  the supersymmetric covariant derivative is defined as: ${\cal D}_{\al} =  \frac{\pa}{\pa \theta^{\al}}\ + \ i \si^{\mu}_{\al \dot{\al}} \, \thb^{\dot{\al}} \, \pa_{\mu} \,\,, \bar{{\cal D}}_{\dot{\al}} =  - \frac{\pa}{\pa \thb^{\dot{\al}}} - \ i \theta^{\al} \, (\si^{\mu})_{\al \dot{\al}} \, \pa_{\mu}$ \cite{Wess&Bagger:1992, MartinP}. Combining the components of $\Phi_+$ with $\Phi_-$ we can construct a 4 component Dirac Spinor ($\psi_D$). 
In the presence of at least 3 flavors of matter fields, the Lagrangian may contain in principle further color singlet terms of dimension $\leq 4$, having the form: $\epsilon_{i j k}(h_+^{f\,f'\,f''}  \Phi^{i\,f}_+ \Phi^{j\,f'}_+ \Phi^{k\,f''}_+ + h_-^{f\,f'\,f''}  \Phi^{i\,f}_- \Phi^{j\,f'}_- \Phi^{k\,f''}_- + h.c.)$, where $h_\pm$ is a totally anti-symmetric tensor with flavor indices; such terms are not included in the present study.

\bigskip
We conclude that the SQCD Lagrangian for ${\cal N}=1$ supersymmetry in 4 dimensions contains, for each flavor of matter fields, two complex scalars (squarks) $A_+ , A_-$, a Dirac spinor (quark) $\{\psi_+ , \psi_-\}$, and two auxiliary complex scalars $F_+, F_-$; in addition, the Lagrangian contains a gauge field (gluon) $u_\mu$, a Majorana spinor (gluino) $\la$ and one further real auxiliary field $D$. Starting from Eq.~(\ref{superfieldLag}), taking the corresponding components of the superfields (appropriate powers of $\theta$ and $\thb$), the continuum Lagrangian in the WZ gauge is: 

\bea
{\cal L}_{\rm SQCD} & = &   - \frac{1}{4}u_{\mu \nu}^{(\alpha)} u^{\mu \nu (\alpha)}
+ \frac{1}{2} D^{(\alpha)} D^{(\alpha)} -i \bar \lambda^{(\alpha)} \bar \sigma^\mu {\cal{D}}_\mu\lambda^{(\alpha)} \nonumber \\
&-& {\cal{D}}_\mu A_+^{\dagger}{\cal{D}}^\mu A_+ - {\cal{D}}_\mu A_-^{\dagger}{\cal{D}}^\mu A_-- i \bar \psi_+ \bar \sigma^\mu {\cal{D}}_\mu \psi_+ - i \bar \psi_- \bar \sigma^\mu {\cal{D}}_\mu \psi_- + F_+^{\dagger} F_+ +F_-^{\dagger} F_-\nonumber \\
&+&i \sqrt2 g  \big( A^{\dagger}_+ \lambda^{(\alpha)} T^{(\alpha)}\psi_+  -  \bar{\psi}_+ \bar \lambda^{(\alpha)}  T^{(\alpha)} A_+ +  A_- \bar{\lambda}^{(\alpha)} T^{(\alpha)} \bar{\psi}_-  - \psi_- \lambda^{(\alpha)}  T^{(\alpha)} A_-\big)\nonumber\\  
&+& g (A^{\dagger}_+ D^{(\alpha)} T^{(\alpha)} A_+ -  A_- D^{(\alpha)} T^{(\alpha)} A^{\dagger}_-)\nonumber \\
&+& m (A_- F_+ + F_- A_+ - \psi_- \psi_+ + A^{\dagger}_+ F^{\dagger}_- + F^{\dagger}_+ A^{\dagger}_- - \bar{\psi}_+ \bar{\psi}_-)\,. 
\label{susylagr1}
\eea

where:

\bea
{\cal{D}}_\mu A_+ &=&  \pa_{\mu} A_+ + i g\,u_{\mu}^{(\alpha)}\,T^{(\alpha)}\,A_+ \nonumber \\
{\cal{D}}_\mu A_-^{\dagger} &=&  \pa_{\mu} A_-^{\dagger} + i g\,u_{\mu}^{(\alpha)}\,T^{(\alpha)}\,A_-^{\dagger} \nonumber \\
{\cal{D}}_\mu A_- &=&  \pa_{\mu} A_- - i g\,A_-\,T^{(\alpha)}\,u_{\mu}^{(\alpha)} \nonumber \\
{\cal{D}}_\mu A_+^{\dagger} &=&  \pa_{\mu} A_+^{\dagger} - i g\,A_+^{\dagger}T^{(\alpha)}\,u_{\mu}^{(\alpha)} \nonumber \\
{\cal{D}}_\mu \psi_+ &=&  \pa_{\mu} \psi_+ + i g\,u_{\mu}^{(\alpha)} \,T^{(\alpha)}\,\psi_+ \nonumber\\
{\cal{D}}_\mu \psi_- &=&  \pa_{\mu} \psi_- - i g \,\psi_- \,T^{(\alpha)}\,u_{\mu}^{(\alpha)} \nonumber \\
{\cal{D}}_\mu \lambda &=&  \pa_{\mu} \lambda + i g \,[u_{\mu},\lambda] \nonumber\\
 u_{\mu \nu} &=& \pa_{\mu}u_{\nu} - \pa_{\nu}u_{\mu} + i g\, [u_{\mu},u_{\nu}]. 
\eea

${\cal L}_{\rm SQCD}$ is invariant, up to a total derivative, under the supersymmetric transformations ($\xi$ is a Majorana spinor parameter):

\bea
\delta_\xi A_+ & = & \sqrt2  \xi \psi_+ \, , \nonumber \\
\delta_\xi A_- & = & \sqrt2  \psi_- \xi  \, , \nonumber \\
\delta_\xi \psi_{+ a} & = & i \sqrt2 \sigma^\mu_{a \dot b} \bar \xi^{\dot b} {\cal{D}}_\mu A_+ + \sqrt2  \xi_a F_+\, , \nonumber \\
\delta_\xi \psi_-^{a} & = & -i \sqrt2 \bar \xi_{\dot b}\bar \sigma^{\dot b a \mu} {\cal{D}}_\mu A_- + \sqrt2 F_- \xi^a\, , \nonumber \\
\delta_\xi F_+ & = & i \sqrt2 \bar \xi \bar \sigma^\mu {\cal{D}}_\mu \psi_+ + 2 i g T^{(\alpha)} A_+ \bar{\xi} \bar{\lambda}^{(\alpha)} \, , \nonumber \\
\delta_\xi F_- & = & -i \sqrt2 {\cal{D}}_\mu \psi_- \sigma^\mu \bar \xi - 2 i g  A_- T^{(\alpha)} \bar{\xi} \bar{\lambda}^{(\alpha)} \, , \nonumber \\
\delta_\xi u_\mu^{(\alpha)} & = & -i \bar \lambda^{(\alpha)} \bar \sigma^\mu \xi + i \bar \xi  \bar \sigma^\mu \lambda^{(\alpha)} \, , \nonumber \\
\delta_\xi \lambda^{(\alpha)} & = & \sigma^{\mu \nu} \xi  u_{\mu \nu}^{(\alpha)}  +i \xi \ D^{(\alpha)}\,, \nonumber \\
\delta_\xi D^{(\alpha)} & = &  - \xi \sigma^\mu {\cal{D}}_\mu \bar{\lambda}^{(\alpha)} - {\cal{D}}_\mu \lambda^{(\alpha)}  \sigma^\mu \bar \xi.
\label{susytransf}
\eea

Note that the above transformations are not linear: Indeed, the standard linear realization of supersymmetry transformations on superfields would reintroduce those field components which are absent in the WZ gauge. Consequently, linear SUSY transformations must be accompanied by appropriate gauge transformations, in order to ensure persistence in the WZ gauge. Thus, the end result is no longer linear in the component fields.

Eq.~(\ref{susylagr1}) can be rewritten in 4 dimensions in Dirac notation and in the Weyl basis as follows:
\bea
{\cal L}_{\rm SQCD} & = &   - \frac{1}{4}u_{\mu \nu}^{\alpha} u^{\mu \nu \alpha}
+ \frac{1}{2} D^{\alpha} D^{\alpha} + \frac{i}{2} \bar \lambda^{\alpha}_M \gamma^\mu {\cal{D}}_\mu\lambda^{\alpha}_M \nonumber \\
&-& {\cal{D}}_\mu A_+^{\dagger}{\cal{D}}^\mu A_+ - {\cal{D}}_\mu A_-{\cal{D}}^\mu A_-^{\dagger}+ i \bar \psi_D \gamma^\mu {\cal{D}}_\mu \psi_D + F_+^{\dagger} F_+ +F_-^{\dagger} F_-\nonumber \\
&-&i \sqrt2 g \big(  A^{\dagger}_+ \bar{\lambda}^{\alpha}_M T^{\alpha} P_+ \psi_D  -  \bar{\psi}_D P_- \lambda^{\alpha}_M  T^{\alpha} A_+ +  A_- \bar{\lambda}^{\alpha}_M T^{\alpha} P_- \psi_D  -  \bar{\psi}_D P_+ \lambda^{\alpha}_M  T^{\alpha} A_-^{\dagger}\big)\nonumber\\  
&+& g (A^{\dagger}_+ D^{\alpha} T^{\alpha} A_+ -  A_- D^{\alpha} T^{\alpha} A^{\dagger}_-)\nonumber \\
&+& m (A_- F_+ + F_- A_+ + \bar \psi_D \psi_D + A^{\dagger}_+ F^{\dagger}_- + F^{\dagger}_+ A^{\dagger}_-),
\eea

\noindent where $P_\pm= \frac{1 \pm \,\gamma_5}{2}$ , $\gamma_5 = i \gamma^0 \gamma^1 \gamma^2 \gamma^3$, $ \la_M= \left( {\begin{array}{c} \la_a\\ \bar \la^{\dot a} \end{array} } \right)$ and $\psi_D^T = \left( {\begin{array}{c} \psi_{+ a}\\ \bar\psi_-^{\dot a} \end{array}} \right)$.

The auxiliary fields may now be eliminated, either by applying their equations of motion (classical case), or by functionally integrating over them (quantum case). In both cases, the action of SQCD takes the following form in Minkowski space:
\bea
{\cal S}_{\rm SQCD} & = & \int d^4x \Big[ -\frac{1}{4}u_{\mu \nu}^{\alpha} {u^{\mu \nu}}^{\alpha} + \frac{i}{2} \bar \lambda^{\alpha}_M \gamma^\mu {\cal{D}}_\mu\lambda^{\alpha}_M \nonumber \\
&-& {\cal{D}}_\mu A_+^{\dagger}{\cal{D}}^\mu A_+ - {\cal{D}}_\mu A_- {\cal{D}}^\mu A_-^{\dagger}+ i \bar \psi_D \gamma^\mu {\cal{D}}_\mu \psi_D \nonumber \\
&-&i \sqrt2 g \big( A^{\dagger}_+ \bar{\lambda}^{\alpha}_M T^{\alpha} P_+ \psi_D  -  \bar{\psi}_D P_- \lambda^{\alpha}_M  T^{\alpha} A_+ +  A_- \bar{\lambda}^{\alpha}_M T^{\alpha} P_- \psi_D  -  \bar{\psi}_D P_+ \lambda^{\alpha}_M  T^{\alpha} A_-^{\dagger}\big)\nonumber\\  
&-& \frac{1}{2} g^2 (A^{\dagger}_+ T^{\alpha} A_+ -  A_- T^{\alpha} A^{\dagger}_-)^2 + m ( \bar \psi_D \psi_D - m A^{\dagger}_+ A_+  - m A_- A^{\dagger}_-)\Big] \,,
\eea

${\cal S}_{\rm SQCD}$ is invariant under supersymmetric transformations:
\bea
\delta_\xi A_+ & = & - \sqrt2  \bar\xi_M P_+\psi_D \, , \nonumber \\
\delta_\xi A_- & = & - \sqrt2  \bar\psi_D P_+ \xi_M  \, , \nonumber \\
\delta_\xi (P_+ \psi_{D}) & = & i \sqrt2 ({\cal{D}}_\mu A_+) P_+ \gamma^\mu \xi_M  - \sqrt2 m P_+ \xi_M A_-^{\dagger}\, , \nonumber \\
\delta_\xi (P_- \psi_D) & = &  i \sqrt2 ({\cal{D}}_\mu A_-)^{\dagger} P_- \gamma^\mu \xi_M  - \sqrt2 m  A_+ P_- \xi_M\, ,\nonumber \\
\delta_\xi u_\mu^{\alpha} & = & -i \bar \xi_M \gamma^\mu \lambda^{\alpha}_M, \nonumber \\
\delta_\xi \lambda^{\alpha}_M & = & \frac{1}{4} u_{\mu \nu}^{\alpha} [\gamma^{\mu},\gamma^{\nu}] \xi_M - 2 i g \gamma^5 \xi_M (A^{\dagger}_+ T^{\alpha} A_+ -  A_- T^{\alpha} A^{\dagger}_-)\,.
\label{susytransfDirac}
\eea

After a Wick rotation, the resulting expression for the Euclidean action in Dirac notation, ${\cal S}^E_{\rm SQCD}$\,, is:\\
\bea
{\cal S}^E_{\rm SQCD} & = & \int d^4x \Big[ \frac{1}{4}u_{\mu \nu}^{\alpha} u_{\mu \nu}^{\alpha} + \frac{1}{2} \bar \lambda^{\alpha}_M \gamma^E_\mu {\cal{D}}_\mu\lambda^{\alpha}_M \nonumber \\
&+& {\cal{D}}_\mu A_+^{\dagger}{\cal{D}}_\mu A_+ + {\cal{D}}_\mu A_- {\cal{D}}_\mu A_-^{\dagger}+ \bar \psi_D \gamma^E_\mu {\cal{D}}_\mu \psi_D \nonumber \\
&+&i \sqrt2 g \big( A^{\dagger}_+ \bar{\lambda}^{\alpha}_M T^{\alpha} P_+^E \psi_D  -  \bar{\psi}_D P_-^E \lambda^{\alpha}_M  T^{\alpha} A_+ +  A_- \bar{\lambda}^{\alpha}_M T^{\alpha} P_-^E \psi_D  -  \bar{\psi}_D P_+^E \lambda^{\alpha}_M  T^{\alpha} A_-^{\dagger}\big)\nonumber\\  
&+& \frac{1}{2} g^2 (A^{\dagger}_+ T^{\alpha} A_+ -  A_- T^{\alpha} A^{\dagger}_-)^2 - m ( \bar \psi_D \psi_D - m A^{\dagger}_+ A_+  - m A_- A^{\dagger}_-)\Big] \,,
\eea
where $P_\pm^E= \frac{1 \pm \,\gamma_5^E}{2}$ , $\gamma_5^E = \gamma_1^E \gamma_2^E \gamma_3^E \gamma_4^E$. Euclidean $\gamma$ matrices are defined as: $\gamma_4^E =\gamma^0$, $\gamma_i^E = -i \gamma_i$  and they satisfy: $\{\gamma_\mu,\gamma_\nu\}=2 \delta_{\mu\,\nu}$.

As is the case with the quantization of ordinary gauge theories, additional infinities will appear upon functionally integrating over gauge orbits. The standard remedy is to introduce a gauge-fixing term in the Lagrangian, along with a compensating Faddeev-Popov ghost term. The resulting Lagrangian, though no longer manifestrly gauge invariant, is still invariant under BRS transformations \cite{Suzuki}. This procedure of gauge fixing guarantees that Green's functions of gauge invariant objects will be gauge independent to all orders in perturbation theory.

For supersymmetric gauge theories one can choose a gauge fixing term \cite{Miller:1983}, which is the natural supersymmetric generalization of covariant gauge fixing (see. Eq.~(\ref{sgf})) :
\bea
S_{GF}^{SUSY}&=& - \frac{1}{8\,\alpha}\int d^4x \left(\bar{{\cal D}}^2V \right)\left({\cal D}^2V \right)|_{\theta \theta \thb \thb}\\\nonumber
&=& - \frac{1}{8\,\alpha\,k} \int d^4x {\rm Tr} \big(4 M \Box M  + 4 N \Box N + 4 (D+\Box C)^2+ 4(\partial_\mu u^\mu)^2 \\\nonumber 
&&-8\la \Box \chi -  8 \bar \la \Box \bar \chi - 8 i \bar \la \bar \si^{\mu} \partial_\mu \la - 8 i \bar \chi \bar \si^{\mu} \partial_\mu \Box \chi\big).
\eea
This gauge fixing term does not break supersymmetry due to the fact that it is a $\theta \theta \thb \thb$ term; thus it is a reasonable choice in regularizations which strive to preserve exact SUSY at all intermediate steps of the calculation of renormalized Green's functions. However, given that the renormalized theory does not depend on the choice of a gauge fixing term, and given that many regularizations, in particular the lattice regularization, violate supersymmetry at intermediate steps, one may as well choose the standard covariant gauge fixing term, proportional to $(\partial_\mu u^\mu)^2$. Actually, this simpler choice is most often utilized also in continuum perturbative calculations of supersymmetric models. Below are the ordinary gauge fixing term and ghost contribution arising from the Faddeev-Popov gauge fixing procedure:
\begin{equation}
{\cal S}^E_{GF}= \frac{1}{\alpha}\int d^4x {\rm{Tr}} \left( \partial_\mu u_\mu\right)^2,
\label{sgf}
\end{equation}
where $\alpha$ is the gauge parameter ($\alpha=1(0)$ corresponds to Feynman (Landau) gauge), and 
\begin{equation}
{\cal S}^E_{Ghost}= - 2 \int d^4x {\rm{Tr}} \left( \bar{c}\, \partial_{\mu}D_\mu  c\right), 
\label{sghost}
\end{equation}
where the ghost field $c$ is a Grassmann scalar which transforms in the adjoint representation of the gauge group, and: ${\cal{D}}_\mu c =  \pa_{\mu} c + i g \,[u_\mu,c]$. The term $S^E_{GF}$ is quadratic with respect to $u_\mu$ and it contributes to the tree level gluon propagator. On the other hand, $S^E_{Ghost}$ contains an interaction between gluon and ghost fields. Consequently, the corresponding continuum action has the form:
\begin{equation}
{\cal S}^E_{\rm total} = {\cal S}^E_{\rm SQCD} + {\cal S}^E_{GF} + {\cal S}^E_{Ghost}.
\label{ScontALL}
\end{equation}

\section{The Calculation of the Supersymmetric Renormalization Factors}
\label{sec3}
In this Section we calculate perturbatively a set of 2-pt and 3-pt Green's functions up to one-loop, both in the continuum and on the lattice. The quantities that we study are the self-energies of the quark ($\psi$), gluon ($u_\mu$), squark ($A$), gluino ($\lambda$), and ghost ($c$) fields, using both dimensional regularization (DR) and lattice regularization (L) \cite{Jack&Jones}. In addition we calculate the gluon-antighost-ghost Green's function in order to renormalize the coupling constant ($g$). The Green's functions leading to self-energies of squarks exhibit also mixing among $A_+$ and $A_-^\dagger$; we calculate the elements of the corresponding $2\times2$ mixing matrix.

Since all of our calculations will be in Euclidean space, the superscript ``$E$'' is understood in what follows.

\subsection{Dimensional Regularization}
\label{sec3.1}

The first step in our perturbative procedure is to calculate the 2-pt and 3-pt Green's functions in the continuum, where we regularize the theory in $D$ Euclidean dimensions ($D=4-2\,\epsilon$). The continuum calculations \cite{Retey, Gracey} are necessary in order to compute the $\MSbar$-renormalized Green's functions; the latter are relevant for the ensuing calculation of the corresponding Green's functions using lattice regularization and $\MSbar$ renormalization. The continuum results also provide the renormalization factors of the quark field ($Z_\psi$), squark field ($Z_{A_\pm}$), gluon field ($Z_u$) gluino field ($Z_\lambda$), ghost field ($Z_c$) and coupling constant ($Z_g$) in DR. For the extraction of the renormalization factors, we applied the $\MSbar$ scheme at a scale $\bar{\mu}$. Once we have computed the renormalization factors in the $\MSbar$ scheme, we can construct their RI$'$ counterparts using conversion factors which are immediately extracted from the above computations to the required perturbative order. Being regularization independent, these same conversion factors can then be also used for lattice renormalization factors.

The aforementioned renormalization factors are defined as follows:
\bea
\psi^R &=& \sqrt{Z_\psi}\,\psi^B,\\
A^R_\pm &=& \sqrt{Z_{A_\pm}}\,A^B_\pm, \\
u_{\mu}^R &=& \sqrt{Z_u}\,u^B_{\mu},\\
\la^R &=& \sqrt{Z_\la}\,\la^B,\\
c^R &=& \sqrt{Z_c}\,c^B, \\
g^R &=& Z_g\,\mu^{-\epsilon}\,g^B, 
\eea
where $B$ stands for bare and $R$ for renormalized quantities, and $\mu$ is an arbitrary scale with dimensions of inverse length. For one-loop calculations, the distinction between $g^R$ and $g^B$ is inessential in many cases; we will simply use $g$ in those cases.

For the calculation of Feynman diagrams in DR we adopt the t'Hooft-Veltman (HV) scheme \cite{Hooft:1972}, in order to continue to $D$ dimensions the metric tensor, $g_D^{\mu\nu}$, and the gamma matrices. The following relations hold in this scheme:
\be
g_D^{\mu\nu}g^D_{\mu\nu}=D,\qquad \{\gamma^\mu,\gamma^\nu\} = 2 g_D^{\mu\nu}.
\ee
The matrix $\gamma_5$ is defined as:
\be
\gamma_5 = \frac{1}{4!}\varepsilon^{\mu\nu\la\rho}\gamma_\mu\gamma_\nu\gamma_\la\gamma_\rho\,,
\ee
where $\varepsilon^{\mu\nu\lambda\rho} = 0$ when any of its indices is outside the range 1-4. In this way:
\be
\{\gamma_5,\gamma_{\mu}\} = 0, \, \,\mu\leq 4,
\ee
\be
[\gamma_5,\gamma_{\mu}]=0, \,\, \mu>4.
\ee

There exist several alternative prescriptions \cite{Larin} for $\gamma_5$\,: Na\"ive dimensional regularization (NDR) \cite{Furman:2003}, in which $\gamma_5$ anticommutes with all $\gamma_\mu$,  as well as the DRED \cite{Siegel:1979} and ${\rm DR}{\overline{{\rm EZ}}}$ schemes (see, e.g., Ref.\cite{Patel}). These prescriptions are related among themselves via finite conversion factors \cite{buras}. Thus, the treatment of diagrams containing quark-squark-gluino interactions in the $\MSbar$ scheme requires special attention. For our continuum results, we use $\MSbar$ renormalization in the HV scheme and for completeness we present also the conversion factors to the NDR scheme. In our calculation of the quark and gluino propagators the indices carried by all gamma matrices are eventually contracted with the indices of external momenta; given that the latter only have 4 (rather than $D$) components, all four  prescriptions give the same results for these propagators. The gluon propagator is also prescription independent to one-loop order, since it does not involve vertices containing $\gamma_5$.

\begin{figure}[h]
\centering
\includegraphics[scale=1]{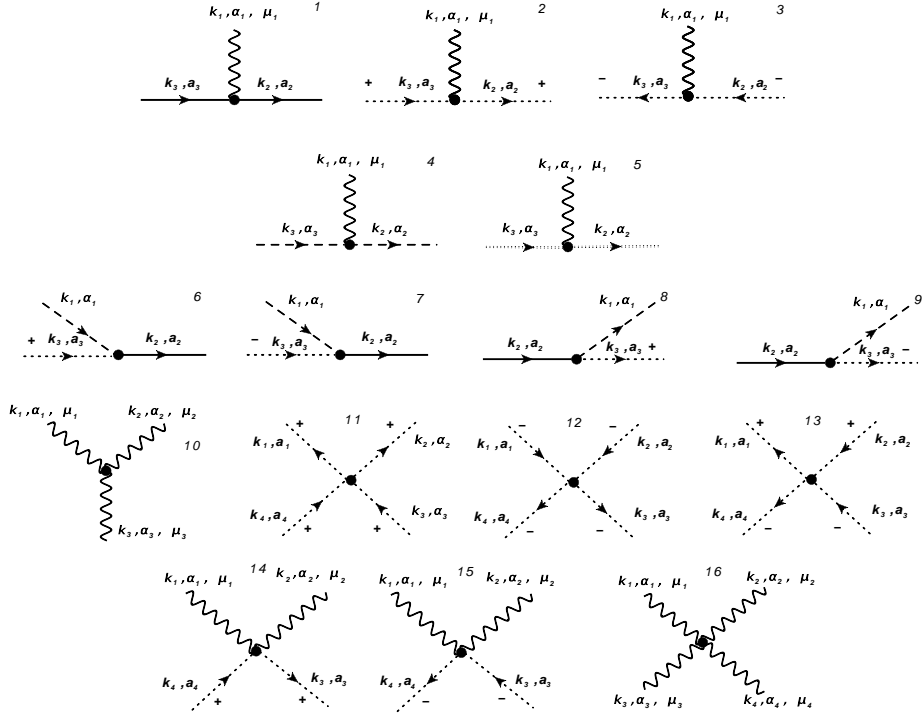}
\caption{Interaction Vertices of ${\cal S}^E_{\rm total}$\,.  A wavy (solid) line represents gluons (quarks). A dotted (dashed) line corresponds to squarks (gluinos). The ``double dashed'' line is the ghost field. Squark lines are further marked with a $+$($-$) sign, to denote an $A_+ \, (A_-)$ field. A squark line arrow entering (exiting) a vertex denotes a $A_+$ ($A_+^{\dagger}$) field; the opposite is true for $A_-$ ($A_-^{\dagger}$) fields.}
\label{vertDR}
\end{figure}

Our conventions for Fourier transformations are: 
\begin{align}
\tilde \psi(q) &= \int d^4x\, e^{-i q \cdot x}\,\psi(x), &  \psi(x) &= \int \frac{d^4q}{(2\pi)^4 } e^{i q \cdot x}\,\tilde \psi(q)\\
\tilde A_\pm(q) &= \int d^4x\, e^{\mp i q \cdot x}\,A_\pm(x), & A_\pm(x) &= \int \frac{d^4q}{(2\pi)^4 } e^{\pm i q \cdot x}\,\tilde A_\pm(q)\\
\tilde{u}_\mu(q) &= \int d^4x\, e^{-i q \cdot x}\,u_\mu(x), & u_\mu(x) &= \int \frac{d^4q}{(2\pi)^4 } e^{i q \cdot x}\,\tilde{u}_\mu(q)\\%
\tilde \la(q) &= \int d^4x\, e^{-i q \cdot x}\,\la(x), & \la(x) &= \int \frac{d^4q}{(2\pi)^4 } e^{i q \cdot x}\,\tilde \la(q)\\
\tilde c(q) &= \int d^4x\, e^{-i q \cdot x}\,c(x), & c(x) &= \int \frac{d^4q}{(2\pi)^4 } e^{i q \cdot x}\,\tilde c(q)
\label{Fourier}
\end{align}
Figure \ref{vertDR} shows all vertices in the action of SQCD (Eq.~(\ref{ScontALL})). There is a total of 16 vertices in the continuum; four of these vertices are present in the non-supersymmetric case ($\#$ 1, 5, 10, 16). The algebraic expression for each vertex, $V_i$ ($i = 1, \ldots, 16$), is given in Eqs.~(\ref{v1}-\ref{v16}); a factor of $\int d^4 k/(2\pi)^4 \tilde X(k)$ is understood for each field $X$ appearing in the vertex; saturation of the vertices' indices (Dirac, color, Lorentz) with those of the corresponding external fields is also implied. [$k_j$ denote momenta; $\alpha_j(a_j)$ are color indices in the adjoint(fundamental) representation; $\mu_j$ are Lorentz indices.]

\bea
\label{v1}
V_1(k_1,k_2,k_3) &=& i g (2\pi)^4 \delta(k_1-k_2+k_3) \gamma_{\mu_1} T^{\alpha_1}_{a_2a_3} {\phantom {1\over 2}}\\
V_2(k_1,k_2,k_3) &=& g (2\pi)^4 \delta(k_1-k_2+k_3)  (k_{2\,\mu_1} + k_{3\,\mu_1})  T^{\alpha_1}_{a_2a_3} {\phantom {1\over 2}}\\
V_3(k_1,k_2,k_3) &=& -g (2\pi)^4 \delta(k_1+k_2-k_3) (k_{2\,\mu_1} + k_{3\,\mu_1}) T^{\alpha_1}_{a_3a_2}{\phantom {1\over 2}}\\
V_4(k_1,k_2,k_3) &=& \frac{1}{2} g (2\pi)^4 \delta(k_1-k_2+k_3)  \gamma_{\mu_1} f^{\alpha_1 \alpha_2 \alpha_3}{\phantom {1\over 2}}\\
V_{5}(k_1,k_2,k_3) &=& - i g(2\pi)^4 \delta(k_1-k_2+k_3) k_{2\,\mu} f^{\alpha_1\,\alpha_2\,\alpha_3} {\phantom {1\over 2}}\\
V_6(k_1,k_2,k_3) &=& - i \sqrt2 g (2\pi)^4 \delta(k_1-k_2+k_3) \frac{1 -  \gamma_5}{2} T^{\alpha_1}_{a_2 a_3}{\phantom {1\over 2}}\\
V_7(k_1,k_2,k_3) &=& - i \sqrt2 g (2\pi)^4 \delta(k_1-k_2+k_3) \frac{1 + \gamma_5}{2} T^{\alpha_1}_{a_2 a_3}{\phantom {1\over 2}}\\
V_8(k_1,k_2,k_3)   &=& i \sqrt2 g (2\pi)^4 \delta(k_1-k_2+k_3)  \frac{1 +  \gamma_5}{2} T^{\alpha_1}_{a_3\,a_2} {\phantom {1\over 2}}\\
V_9(k_1,k_2,k_3) &=& i \sqrt2 g (2\pi)^4 \delta(k_1-k_2+k_3) \frac{1 -  \gamma_5}{2} T^{\alpha_1}_{a_3 a_2}{\phantom {1\over 2}}\\
V_{10}(k_1,k_2,k_3) &=& - \frac{i}{2} g (2\pi)^4 \delta(k_1+k_2+k_3) f^{\alpha_1\,\alpha_2\,\alpha_3}\delta_{\mu_1 \mu_2} (k_{2\,\mu_3}-k_{1\,\mu_3}) {\phantom {1\over 2}}\\
V_{11}(k_1,k_2,k_3,k_4) &=& \frac{1}{2} g^2 (2\pi)^4 \delta(k_1+k_2-k_3-k_4) T^{\alpha}_{a_1\,a_3} T^{\alpha}_{a_2\,a_4}{\phantom {1\over 2}}\\
V_{12}(k_1,k_2,k_3,k_4) &=& \frac{1}{2} g^2 (2\pi)^4 \delta(k_1+k_2-k_3-k_4) T^{\alpha}_{a_3\,a_1} T^{\alpha}_{a_4\,a_2}{\phantom {1\over 2}}\\
V_{13}(k_1,k_2,k_3,k_4) &=& - g^2 (2\pi)^4 \delta(k_1-k_2+k_3-k_4) T^{\alpha}_{a_1\,a_2} T^{\alpha}_{a_4\,a_3}{\phantom {1\over 2}}\\
V_{14}(k_1,k_2,k_3,k_4) &=& g^2 (2\pi)^4 \delta(k_1+k_2-k_3+k_4) (T^{\alpha_1} T^{\alpha_2})_{a_3\,a_4} \delta_{\mu_1\,\mu_2}{\phantom {1\over 2}}\\
V_{15}(k_1,k_2,k_3,k_4) &=& g^2 (2\pi)^4 \delta(k_1+k_2+k_3-k_4) (T^{\alpha_1} T^{\alpha_2})_{a_4\,a_3} \delta_{\mu_1\,\mu_2}{\phantom {1\over 2}}\\
V_{16}(k_1,k_2,k_3,k_4) &=& \frac{1}{4} g^2 (2\pi)^4 \delta(k_1+k_2+k_3+k_4)f^{\alpha_1\,\alpha_2\,\alpha}f^{\alpha\,\alpha_3\,\alpha_4} \delta_{\mu_1\,\mu_3}\delta_{\mu_2\,\mu_4}
\label{v16}
\eea
All these vertices are intended to be symmetrized over identical fields before contraction among the fields and creation of Feynman diagrams; a summation over the color index $\alpha$ is understood.

\begin{figure}[ht!]
\centering
\includegraphics[scale=0.55]{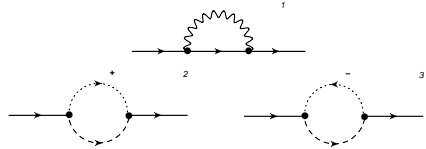}
\caption{One-loop Feynman diagrams contributing to the 2-pt Green's function $\langle \psi(x) \bar \psi(y) \rangle$.
  }
\label{quark2pt}
\end{figure}

\begin{figure}[ht!]
\centering
\includegraphics[scale=0.75]{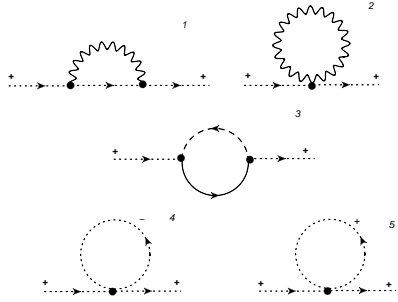}
\caption{One-loop Feynman diagrams contributing to the 2-pt Green's function 
$\langle A_+(x) A_+^{\dagger}(y) \rangle$. The case of $\langle  A_-^{\dagger}(x) A_-(y) \rangle$ is completely analogous.
  }
\label{squark2pt}
\end{figure}

\begin{figure}[ht!]
\centering
\includegraphics[scale=0.75]{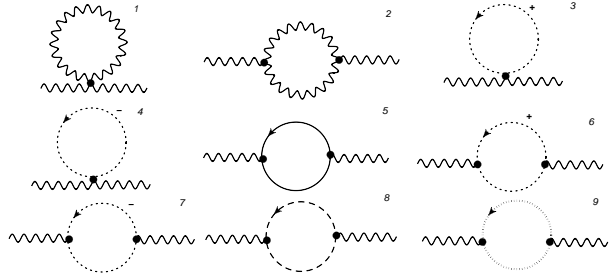}
\caption{One-loop Feynman diagrams contributing to the 2-pt Green's function  $\langle  u_\mu^{\alpha}(x) u_\nu^{\beta}(y) \rangle$. 
  }
\label{gluon2pt}
\end{figure}

\begin{figure}[ht!]
\centering
\includegraphics[scale=0.55]{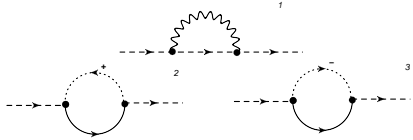}
\caption{One-loop Feynman diagrams contributing to the 2-pt Green's function  $\langle  \lambda^\alpha(x) \bar \lambda^\beta(y) \rangle$.
  }
\label{gluino2pt}
\end{figure}

\begin{figure}[ht!]
\centering
\includegraphics[scale=0.2]{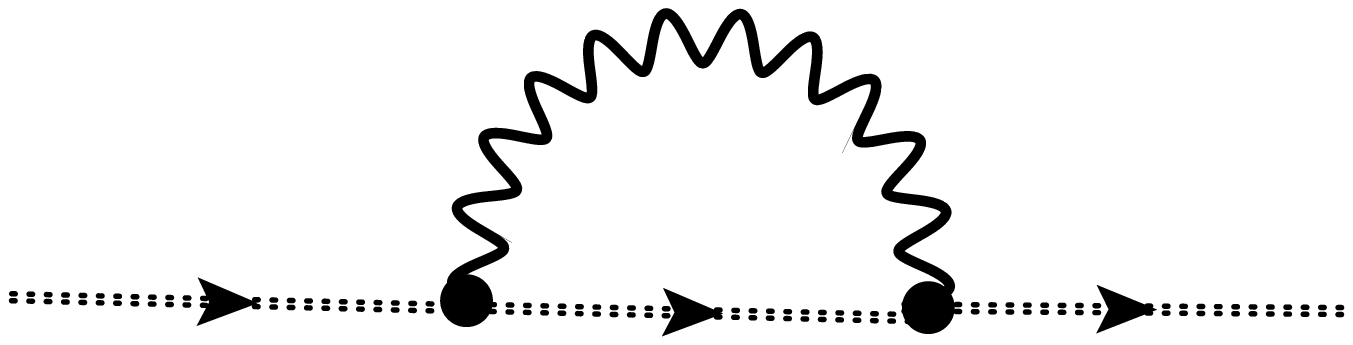}
\caption{One-loop Feynman diagram contributing to the 2-pt Green's function $\langle  c(x) \bar c(y) \rangle$.
  }
\label{ghost2pt}
\end{figure}

The one-loop Feynman diagrams (one-particle irreducible (1PI)) contributing to the quark propagator, $\langle \psi(x) \bar \psi(y) \rangle$,  are shown in Fig.~\ref{quark2pt}, those contributing to the squark propagator, $\langle A_+(x) A^{\dagger}_+(y) \rangle$, in Fig.~\ref{squark2pt}. Identical results are obtained for $\langle A_+(x) A_+^{\dagger}(y) \rangle$ and $\langle  A_-^{\dagger}(x) A_-(y)\rangle$. The one-loop Feynman diagrams contributing to the gluon propagator, $\langle  u_\mu^{\alpha}(x) u_\nu^{\beta}(y) \rangle$, and gluino propagator, $\langle  \lambda^{\alpha}(x) \bar \lambda^{\beta}(y) \rangle$, are shown in Fig.~\ref{gluon2pt} and Fig.~\ref{gluino2pt}, respectively. Lastly, the 1PI Feynman diagram which contributes to the ghost propagator, $\langle c(x) \bar c(y) \rangle$, is shown in Fig.~\ref{ghost2pt}. As is usually done, we will work in a mass-independent scheme, and thus all of our calculations, in the continuum as well as on the lattice will be done at zero renormalized masses for all particles.

Here we collect all our results for the 2-pt inverse Green's functions; the first result which we present, is the inverse quark propagator in momentum space:
\be
\langle \tilde \psi^B(q) \tilde{\bar{\psi}}^B(q') \rangle^{DR}_{\rm{inv}} = (2\pi)^4 \delta(q-q') i \qslash \left[ 1 + \frac{g^2\,C_F}{16\,\pi^2} \left( \frac{2+\alpha}{\epsilon} + 4 + \alpha +  (2+\alpha) \log\left(\frac{\bar\mu^2}{q^2} \right) \right) \right],
\label{GF2quark}
\ee
where $C_F=(N_c^2-1)/(2\,N_c)$ is the quadratic Casimir operator in the fundamental representation, $q$ is the external momentum in the Feynman diagrams, and $\bar\mu$ is the energy scale which is related to $\mu$ through\footnote{$\gamma_E$  is Euler's constant: $\gamma_E = 0.57721$\, .}: $\mu = \bar \mu \sqrt{e^{\gamma_E}/ 4\pi}$. Note also that a Kronecker delta for color indices is understood in Eqs.~(\ref{GF2quark})-(\ref{GF2ghost}).

In the HV prescription, there is mixing in the 2-pt Green's functions of $A_+$ and $A_-$. The diagonal elements of the mixing matrix are: 
\be
\langle \tilde A_+^B(q) \tilde A_+^{B\,\dagger}(q') \rangle^{DR}_{\rm{inv}} =\langle \tilde A_-^{B\,\dagger}(q) \tilde A_-^B(q') \rangle^{DR}_{\rm{inv}} = (2\pi)^4 \delta(q-q') q^2 \left[ 1 + \frac{g^2\,C_F}{16\,\pi^2} \left( \frac{1+\alpha}{\epsilon} + \frac{16}{3} +  (1+\alpha) \log\left(\frac{\bar\mu^2}{q^2} \right) \right) \right].
\label{GF2squark}
\ee
The nondiagonal elements are shown in Eq.~(\ref{GF2squarkmixed}). Furthermore, we calculate here the same quantities in the NDR scheme. It is known that this scheme needs additional corrections \cite{HP}, in order, e.g., to reproduce the correct axial anomaly in QCD. The diagonal elements in NDR are:
\be
\langle \tilde A_+^B(q) \tilde A_+^{B\,\dagger}(q') \rangle^{NDR}_{\rm{inv}} =\langle \tilde A_-^{B\,\dagger}(q) \tilde A_-^B(q') \rangle^{NDR}_{\rm{inv}} = (2\pi)^4 \delta(q-q') q^2 \left[ 1 + \frac{g^2\,C_F}{16\,\pi^2} \left( \frac{1+\alpha}{\epsilon} + 4 +  (1+\alpha) \log\left(\frac{\bar\mu^2}{q^2} \right) \right) \right].
\label{GF2squarkDRED}
\ee
Note that, because of the definition of $\gamma_5$ in the NDR prescription, the nondiagonal elements vanish. In Eq.~(\ref{conversiontoNDR}), we have determined the conversion factor between NDR and HV.

We now turn to the gluon propagator. The contributions from the diagrams of Fig.~\ref{gluon2pt}, taken separately, are not transverse. But, their sum has this property, and it is found to take the following form in the continuum:
\bea
\langle  \tilde u_\mu^{B}(q) \tilde u_\nu^{B}(q') \rangle^{DR}_{\rm{inv}} &=&(2\pi)^4 \delta(q+q') \Bigg\{  \frac{1}{\alpha} q_{\mu} q_{\nu}\nonumber\\\nonumber
&& + \left(q^2 \delta_{\mu \nu} - q_{\mu} q_{\nu}\right)\Bigg[ 1  + \frac{g^2\, N_f}{16\,\pi^2} \left( \frac{1}{\epsilon} + 2 +\log\left(\frac{\bar\mu^2}{q^2} \right)\right)\\\nonumber
&& - \frac{g^2\, N_c}{16\,\pi^2}\frac{1}{2} \left(\left(3-\alpha\right)\frac{1}{\epsilon} + \frac{19}{6} + \alpha + \frac{\alpha^2}{2} + \left(3-\alpha\right)\log\left(\frac{\bar\mu^2}{q^2} \right)\right)\Bigg] \Bigg\}\\
\eea
Since there is no one-loop longitudinal part for the gluon self-energy, the renormalization factor for the gauge parameter receives no one-loop contribution. This result as well as the result for the quark propagator are in complete agreement with older results in the absence of supersymmetry (see, e.g., Ref.~\cite{Gracey:2003}).

Our result for the inverse gluino propagator to one-loop order is:
\bea
\langle \tilde \lambda^{B}(q) \tilde{\bar{\lambda}}^{B}(q') \rangle^{DR}_{\rm{inv}} &=& (2\pi)^4 \delta(q-q') \frac{i}{2}\, \qslash \Bigg[ 1 + \frac{g^2\,N_f}{16\,\pi^2}\left( 2 + \frac{1}{\epsilon} + \log\left(\frac{\bar\mu^2}{q^2} \right)\right)\\\nonumber
&& + \frac{g^2\, N_c}{16\,\pi^2} \left(\alpha + \frac{\alpha}{\epsilon} + \alpha \log\left(\frac{\bar\mu^2}{q^2}  \right)\right)\Bigg].
\eea

Finally, the ghost propagator is the same as in the non-supersymmetric case, since its contributions come only from $S_{Ghost}$:
\be
\langle \tilde c^{B}(q) \tilde{\bar{c}}^{B}(q') \rangle^{DR}_{\rm{inv}} = (2\pi)^4 \delta(q-q') q^2 \left[ 1 - \frac{g^2\,N_c}{16\,\pi^2} \left(1 + \frac{3-\alpha}{4\epsilon}  + \frac{1}{4} (3-\alpha) \log\left(\frac{\bar\mu^2}{q^2} \right) \right) \right].
\label{GF2ghost}
\ee

Starting from the 2-pt Green's functions, it is straightforward to write down the renormalization factors for all fields appearing in ${\cal S}^E_{\rm total}$ in the $\MSbar$ renormalization scheme. In this scheme, renormalization factors are simply defined in such a way as to only remove the pole parts.

The results for the DR renormalization factors in the $\MSbar$ scheme are presented for arbitrary values of $N_c$, $N_f$, and  $\alpha$:
\bea
Z_\psi^{DR,\MSbar} &=& 1 + \frac{g^2\,C_F}{16\,\pi^2} \frac{1}{\epsilon}\left( 2 + \alpha \right)\\
Z_{A_\pm}^{DR,\MSbar} &=& 1 + \frac{g^2\,C_F}{16\,\pi^2} \frac{1}{\epsilon}\left(1 + \alpha \right)\\
Z_{u}^{DR,\MSbar} &=&  1 + \frac{g^2\,}{16\,\pi^2} \frac{1}{\epsilon} \left[\left(\frac{\alpha}{2} - \frac{3}{2}\right) N_c +  N_f \right]\\
Z_{\lambda}^{DR,\MSbar} &=&  1 + \frac{g^2\,}{16\,\pi^2} \frac{1}{\epsilon} \left(\alpha\, N_c + N_f \right)\\
Z_{c}^{DR,\MSbar} &=& 1 - \frac{g^2\,}{16\,\pi^2} \frac{1}{\epsilon} \frac{3-\alpha}{4}
\eea

\begin{figure}[ht!]
\centering
\includegraphics[scale=0.4]{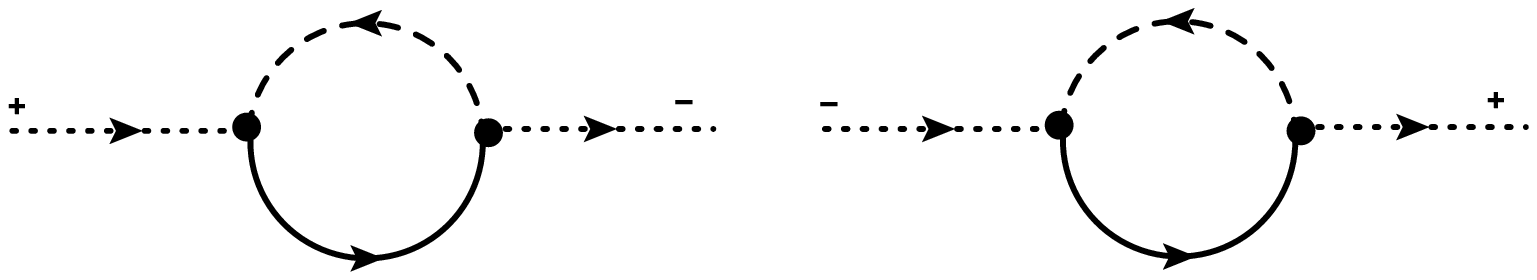}
\caption{One-loop Feynman diagrams contributing to the 2-pt Green's function 
$\langle \tilde A_+^B(q) \tilde A_-^{B}(q') \rangle$ and $\langle \tilde A_-^{B\,\dagger}(q) \tilde A_+^{B\,\dagger}(q') \rangle$.
  }
\label{mixed2pt}
\end{figure}

As already mentioned, there are also nonzero 2-pt Green's functions connecting $A_+$ with $A_-$ in the HV scheme. The mixed Green's functions of $A_+$ and $A_-$ arise beyond tree level: $\langle \tilde A_+^B(q) \tilde A_-^{B}(q') \rangle$ and $\langle \tilde A_-^{B\,\dagger}(q) \tilde A_+^{B\,\dagger}(q') \rangle$. This nonsingular mixing comes from the diagrams of Fig.~\ref{mixed2pt} and the corresponding mixing elements obtained in the HV scheme are shown in Eq.~(\ref{GF2squarkmixed}).  
Given that the $\MSbar$ scheme affects only the pole parts ($1/\epsilon$ terms), the renormalization factors in this scheme will be diagonal. However, the 2-pt Green's functions of the $\MSbar$-renormalized squark fields will remain mixed. 

The nondiagonal elements are equal to each other and our result, to one loop, is:
\be
\langle \tilde A_+^B(q) \tilde A_-^{B}(q') \rangle^{DR}_{\rm{inv}}=\langle \tilde A_-^{B\,\dagger}(q')\tilde A_+^{B\,\dagger}(q') \rangle^{DR}_{\rm{inv}} = (2\pi)^4 \delta(q-q') \frac{g^2\,C_F}{16\,\pi^2}  \frac{4}{3} q^2. 
\label{GF2squarkmixed}
\ee

In matrix notation, our results for the $\MSbar$ renormalized Green's functions are:
\bea
\label{MIXmatrix}
\hspace{-0.5cm}\langle \tilde A^\MSbar(q) {\tilde A}^{\MSbar\,\dagger}(q') \rangle_{\rm{inv}}&=& (2\pi)^4 \delta(q-q')\Bigg[ q^2 \begin{pmatrix} 1 & 0\\ 0 & 1 \end{pmatrix} + q^2 \frac{g^2\,C_F}{16\,\pi^2} \left( \frac{16}{3} +  (1+\alpha) \log\left(\frac{\bar\mu^2}{q^2}\right)\right)\begin{pmatrix} 1 & 0 \\ 0 & 1 \end{pmatrix} + q^2 \frac{g^2\,C_F}{16\,\pi^2}\frac{4}{3}\begin{pmatrix} 0 & 1\\ 1 & 0 \end{pmatrix}\Bigg]\nonumber\\
&\equiv&(2\pi)^4 \delta(q-q')\left[ q^2 \openone - \Sigma^{\MSbar} \right].
\eea
where $A^\MSbar$ is a 2-component column which contains the renormalized squark fields:
\be
A^\MSbar = \left( {\begin{array}{c} A^\MSbar_+ \\ {A_-^\MSbar}^\dagger \end{array} } \right)
\ee

\medskip
Since the nondiagonal matrix elements of $\langle A^\MSbar A^{\MSbar\,\dagger} \rangle_{\rm{inv}}$ are simply constant multiples of $q^2$, we can use another renormalization scheme ($Y$), in which the squark field propagators are disentangled. The two schemes will be related via a mixing matrix (conversion factor), $C^{Y,\,\MSbar}$, which is finite and scale independent. The conversion factor does not depend on the regularization scheme (thus, it will be the same also using a lattice regularization, as in Section~\ref{sec3.2}) and its definition is as follows: 

\be
A^Y \equiv \left( C^{Y,\MSbar} \right)^{1/2}\,  A^\MSbar.
\label{Diag1}
\ee

There are several possible choices for such a scheme:  

\begin{itemize}
\item A ``diagonal'' scheme (``$D$'') in which the diagonal elements of the squark propagator matrix remain unaffected. 
\be 
C^{D,\,\MSbar} = \openone+ \frac{g^2\,C_F}{16\,\pi^2} \frac{4}{3}\begin{pmatrix} 0 & 1\\ 1 & 0 \end{pmatrix}
\label{conversiontoD}
\ee
\item An RI$'$-like scheme (``RI$'$''), in which the renormalization prescription is: $\Sigma^{\rm RI'}|_{q^2 = \bar\mu^2} = 0$.
\be C^{{\rm RI'},\, \MSbar} = \openone+ \frac{g^2\,C_F}{16\,\pi^2} \frac{4}{3}\begin{pmatrix} 4 & 1\\ 1 & 4 \end{pmatrix}
\label{conversiontoRI}
\ee
\item An $\MSbar$-like scheme (``$\MSbar$NDR''), in which bare Green's functions are constructed in the NDR regularization, and only pole parts
in $\epsilon$ are removed from them.
\be C^{\MSbar{\rm NDR},\, \MSbar} = \openone+ \frac{g^2\,C_F}{16\,\pi^2} \frac{4}{3}\begin{pmatrix} 1 & 1\\ 1 & 1 \end{pmatrix}
\label{conversiontoNDR}
\ee
\end{itemize}

\begin{figure}[htb!]
\centering
\includegraphics[scale=0.75]{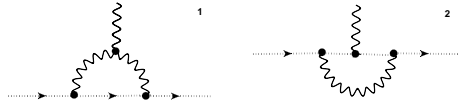}
\caption{One-loop Feynman diagrams contributing to $\langle c^{\alpha}(x) \bar{c}^{\beta}(y) u_\mu^{\gamma}(z)\rangle$.  
}
\label{3ptDR}
\end{figure}

In this paper we also calculate the gluon-antighost-ghost Green's function in order to renormalize the coupling constant. Other determinations of the coupling constant renormalization (through gluon-antiquark-quark, gluon-gluino-gluino and gluon-antisquark-squark Green's functions) are expected to lead to identical results. We compute, perturbatively to one-loop, this Green's function using both dimensional and lattice regularizations.  In Fig.~\ref{3ptDR}, we have drawn the corresponding continuum 1PI Feynman diagrams.
The 3-pt amputated Green's function, at zero antighost momentum, in DR gives:
\be
\langle \tilde c^{B\,\alpha}(q) \tilde{\bar{c}}^{B\,\beta}(0) \tilde{u}_\mu^{B\,\gamma}(q')\rangle^{DR}_{\rm{amp}}  = 
(2\pi)^4 \delta(q+q') f^{\alpha\,\beta\,\gamma} (i g q_\mu) \Big[1 + \frac{g^2\,N_c}{16\,\pi^2}\frac{\alpha}{2} \Big(1 + \frac{1}{\epsilon}  +  \log \left(\frac{\bar{\mu}^2}{q^2} \right)\Big].
\label{GF3ghostsgluon}
\ee
The value of $Z_g^{DR,\MSbar}$ now stems from the requirement:
\be
\lim_{\epsilon \to 0}\Big[\left(Z_{c}^{DR,\MSbar}\right)^{-1} \left(Z_{u}^{DR,\MSbar}\right)^{-1/2}\left(Z_{g}^{DR,\MSbar}\right)^{-1}  \langle \tilde c^{B\,\alpha}(q) \tilde{\bar{c}}^{B\,\beta}(0) \tilde{u}_\mu^{B\,\gamma}(q')\rangle^{DR}_{\rm{amp}} \Big] = \langle \tilde c^{B\,\alpha}(q) \tilde{\bar{c}}^{B\,\beta}(0) u_\mu^{B\,\gamma}(q')\rangle^{DR}_{\rm{amp}}\Big{|}_{1/\epsilon \to 0}
\label{GF3condition}
\ee
In the above equation $Z_g^{DR,\MSbar}$ is required to eliminate only the pole parts of the left-hand side, without additional finite terms; the same requirement leads to the definition of $Z_g^{DR,\rm RI'}$, which is thus trivially equal to $Z_g^{DR,\MSbar}$. The right-hand side is actually the $\MSbar$ renormalized 3-pt Green's function. There follows:
\be
Z_{g}^{DR,\MSbar}=1 + \frac{g^2\,}{16\,\pi^2} \frac{1}{\epsilon} \left(\frac{3}{2} N_c - \frac{1}{2}N_f \right).\\
\ee
Notice that the expression for $Z_{g}^{DR,\MSbar}$ is gauge independent.

\subsection{Lattice Regularization}
\label{sec3.2}

Even though the lattice breaks supersymmetry explicitly \cite{Feo}, it is the only regulator which describes many aspects of strong interactions nonperturbatively. We use a standard discretization where the quarks, squarks and gluinos live on the lattice sites and the gluons live on the links of the lattice: $U_\mu (x) = e^{i g a T^{\alpha} u_\mu^\alpha (x+a\hat{\mu}/2)}$. We will extend Wilson's formulation of the QCD action, to encompass SUSY partner fields as well. This formulation leaves no SUSY generators intact, and it also breaks chiral symmetry; it thus represents a ``worst case'' scenario, which is worth investigating in order to address the complications which will arise in numerical simulations of SUSY theories. In our ongoing investigation we plan to address also improved actions, so that we can check to what extent some of the SUSY breaking effects can be alleviated.

\begin{figure}[ht!]
\centering
\includegraphics[scale=0.9]{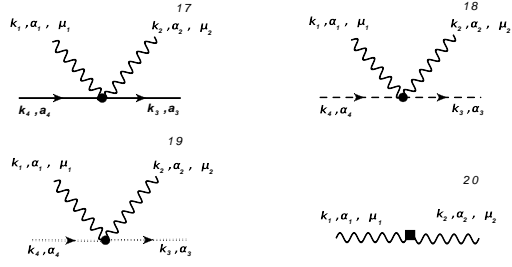}
\caption{Additional interaction vertices in Lattice-SQCD. All fields are represented as in Fig.~\ref{vertDR}. The solid box in the bottom right vertex comes from the measure part of the lattice action.
  }
\label{EXTRAlattice}
\end{figure}

As we mentioned earlier, we will calculate the renormalization factors which are necessary ingredients in relating lattice matrix elements to physical amplitudes. Our computation is performed to one loop order and to the lowest order in the lattice spacing, $a$. Lattice perturbation theory is much more complicated than continuum perturbation theory. There are more vertices stemming from the discretized action, see Fig.~\ref{EXTRAlattice}, leading to  more Feynman diagrams; what is worse, the propagators and vertices, with which one builds the Feynman diagrams, are also more complicated on the lattice than they are in the continuum, which can lead to expressions containing a very large number of terms: Even in the Wilson formulation, which is rather concise, a typical ``difficult'' Green's function contains a few thousand terms at intermediate stages.

\begin{figure}[ht!]
\centering
\includegraphics[scale=0.75]{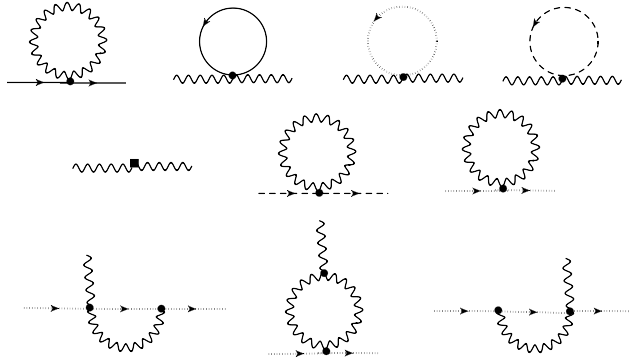}
\caption{Additional one-loop Feynman diagrams contributing to the 2-pt and 3-pt Green's functions on the lattice. 
  }
\label{EXTRAlattice1}
\end{figure}

Calculating the same Green's functions as before on the lattice, and combining them with our results from the continuum, we will be able to extract $Z_\psi^{L,\MSbar}$, $Z_u^{L,\MSbar}$, $Z_\lambda^{L,\MSbar}$, $Z_{A_\pm}^{L,\MSbar}$, $Z_c^{L,\MSbar}$ and $Z_g^{L,\MSbar}$ in the $\MSbar$ scheme and on the lattice. On the lattice we have to calculate all the diagrams which were presented in Sec.~\ref{sec3.1} as well as further loop diagrams, as shown in Fig.~\ref{EXTRAlattice1}; in addition, for the gluon propagator we have to take into account the contribution which comes from the measure part of the action. For the algebraic operations involved in evaluating Feynman diagrams, we make use of our symbolic package in Mathematica. 

For Wilson-type fermions and gluons, the Euclidean action ${\cal S}^{L}_{\rm SQCD}$ on the lattice becomes:          
\bea
{\cal S}^{L}_{\rm SQCD} & = & a^4 \sum_x \Big[ \frac{N_c}{g^2} \sum_{\mu,\,\nu}\left(1-\frac{1}{N_c}\, {\rm Tr} U_{\mu\,\nu} \right ) + \sum_{\mu} {\rm Tr} \left(\bar \lambda_M \gamma_\mu {\cal{D}}_\mu\lambda_M \right ) - a \frac{r}{2} {\rm Tr}\left(\bar \lambda_M  {\cal{D}}^2 \lambda_M \right) \nonumber \\ 
&+&\sum_{\mu}\left( {\cal{D}}_\mu A_+^{\dagger}{\cal{D}}_\mu A_+ + {\cal{D}}_\mu A_- {\cal{D}}_\mu A_-^{\dagger}+ \bar \psi_D \gamma_\mu {\cal{D}}_\mu \psi_D \right) - a \frac{r}{2} \bar \psi_D  {\cal{D}}^2 \psi_D \nonumber \\
&+&i \sqrt2 g \big( A^{\dagger}_+ \bar{\lambda}^{\alpha}_M T^{\alpha} P_+ \psi_D  -  \bar{\psi}_D P_- \lambda^{\alpha}_M  T^{\alpha} A_+ +  A_- \bar{\lambda}^{\alpha}_M T^{\alpha} P_- \psi_D  -  \bar{\psi}_D P_+ \lambda^{\alpha}_M  T^{\alpha} A_-^{\dagger}\big)\nonumber\\  
&+& \frac{1}{2} g^2 (A^{\dagger}_+ T^{\alpha} A_+ -  A_- T^{\alpha} A^{\dagger}_-)^2 - m ( \bar \psi_D \psi_D - m A^{\dagger}_+ A_+  - m A_- A^{\dagger}_-)\Big] \,,
\label{susylagrLattice}
\eea
where: $U_{\mu \nu}(x) =U_\mu(x)U_\nu(x+a\hat\mu)U^\dagger_\mu(x+a\hat\nu)U_\nu^\dagger(x)$, and a summation over flavors is understood in the last three lines of Eq.~(\ref{susylagrLattice}). The 4-vector $x$ is restricted to the values $x = na$, with $n$ being an integer 4-vector.
Thus the momentum integration, after a Fourier transformation, is restricted to the first Brillouin zone (BZ) $[ - \pi/a,\pi/a ]^4$ and the sum over $x$ leads to momentum conservation in each vertex. 
The terms proportional to the Wilson parameter, $r$, eliminate the problem of fermion doubling, at the expense of breaking chiral invariance~\footnote{In what follows, we will set $|r|=1$.}.

The definitions of the covariant derivatives are as follows:
\bea
{\cal{D}}_\mu\lambda_M(x) &\equiv& \frac{1}{2a} \Big[ U_\mu (x) \lambda_M (x + a \hat{\mu}) U_\mu^\dagger (x) - U_\mu^\dagger (x - a \hat{\mu}) \lambda_M (x - a \hat{\mu}) U_\mu(x - a \hat{\mu}) \Big] \\
{\cal D}^2 \lambda_M(x) &\equiv& \frac{1}{a^2} \sum_\mu \Big[ U_\mu (x)  \lambda_M (x + a \hat{\mu}) U_\mu^\dagger (x)  - 2 \lambda_M(x) +  U_\mu^\dagger (x - a \hat{\mu}) \lambda_M (x - a \hat{\mu}) U_\mu(x - a \hat{\mu})\Big]\\
{\cal{D}}_\mu \psi_D(x) &\equiv& \frac{1}{2a}\Big[ U_\mu (x) \psi_D (x + a \hat{\mu})  - U_\mu^\dagger (x - a \hat{\mu}) \psi_D (x - a \hat{\mu})\Big]\\  
{\cal D}^2 \psi_D(x) &\equiv& \frac{1}{a^2} \sum_\mu \Big[U_\mu (x) \psi_D (x + a \hat{\mu})  - 2 \psi_D(x) +  U_\mu^\dagger (x - a \hat{\mu}) \psi_D (x - a \hat{\mu})\Big]\\
{\cal{D}}_\mu A_+(x) &\equiv& \frac{1}{a} \Big[  U_\mu (x) A_+(x + a \hat{\mu}) - A_+(x)   \Big]\\
{\cal{D}}_\mu A_+^{\dagger}(x) &\equiv& \frac{1}{a} \Big[A_+^{\dagger}(x + a \hat{\mu}) U_\mu^{\dagger}(x)  -  A_+^\dagger(x)\Big]\\
{\cal{D}}_\mu A_-(x) &\equiv& \frac{1}{a} \Big[A_-(x + a \hat{\mu}) U_\mu^{\dagger}(x)  -  A_-(x)\Big]\\
{\cal{D}}_\mu A_-^{\dagger}(x) &\equiv& \frac{1}{a} \Big[U_\mu (x) A_-^{\dagger}(x + a \hat{\mu})   -  A_-^{\dagger}(x) \Big]
\eea

As always in perturbation theory, we must introduce an appropriate gauge-fixing term to the action; in terms of the gauge field $u_\mu(x)$ it reads:
\be
S_{GF}^{L} = \frac{1}{2\alpha} a^2 \sum_x \sum_\mu{\rm{Tr}} \left(u_\mu(x + a \hat{\mu}/2)-u_\mu(x - a \hat{\mu}/2)\right)^2.
\ee
In simulations there is no need for gauge fixing since functional integration is performed over a finite number of degrees of freedom (d.o.f.), each of which ranges within the compact domain of the group manifold. However, in perturbation theory, where an infinite number of d.o.f. takes values over the noncompact algebra, gauge fixing is necessary in order to avoid divergences from the integration over gauge orbits. Covariant gauge fixing produces the following action for the ghost fields $c$ and $\bar c$: 
\bea
S_{Ghost}^{L} &=& 2a^2 \sum_x\sum_\mu {\rm{Tr}} \{(\bar c(x + a \hat{\mu}) - \bar c(x))( c(x + a \hat{\mu}) - c(x) \\\nonumber
&&+ ig [u_\mu (x+ a \hat{\mu}/2),c(x)] + \frac{1}{2}ig [u_\mu (x+ a \hat{\mu}/2),c(x+a \hat{\mu}) -c(x)]\\\nonumber
&&-\frac{1}{12}g^2[u_\mu (x+ a \hat{\mu}/2),[u_\mu (x+ a \hat{\mu}/2),c(x+a \hat{\mu}) -c(x)]])\}+{\cal{O}}(g^3).
\eea
In Eq.~(\ref{susylagrLattice}) we must also add a term arising from the fact that we change integration variables from the link variables $U_\mu$ to the gluon fields $u_\mu$\,. This procedure changes the Haar integration measure by a Jacobian, which can be recast as an exponential, thus taking the form of an additional contribution to the action; this is the usual measure term $S_{M}^{L}$ in the action and, to lowest order in $g$, it reads:
\be
S_{M}^{L} = \frac{g^2N_c}{12}a^2\sum_x \sum_\mu {\rm{Tr}}\left(u_\mu(x+ a \hat{\mu}/2)\right)^2 + {\cal{O}}(g^4).
\ee
The terms ${\cal S}^{L}_{\rm SQCD}$, $S_{GF}^{L}$, $S_{Ghost}^{L}$ and $S_{M}^{L}$ must be added to obtain the total lattice action.

We now summarize the rules for calculating the contribution of a Feynman diagram to the 2-pt and 3-pt Green's functions. In the following equations we present the tree-level propagators.

\be
\begin{array}{ll}
{\rm Quark\ propagator:} \qquad & {\displaystyle \frac{1}{i \,\qcircslash + {2r\over a} \sum_\mu \sin^2(aq_\mu/2) - m}},\qquad{\rm where:}\qquad {\displaystyle \qcircslash = {1\over a}\,\sum_\mu\gamma_\mu  \,\sin(aq_\mu)} \\\\
{\rm Gluon\ Propagator:}\qquad & {\displaystyle \frac{1}{\hat{q}^2} \left(\delta_{\mu\nu} - (1-\alpha)\frac{\hat{q}_\mu\hat{q}_\nu}{\hat{q}^2} \right)}, \qquad{\rm where:}\qquad {\displaystyle \hat{q}_\mu = \frac{2}{a}\sin\frac{aq_\mu}{2}\,, \quad \hat{q}^2 = \sum_\mu \hat{q}_\mu^2} \\\\
{\rm Ghost\ propagator:} \qquad & {\displaystyle \frac{1}{\hat{q}^2}} \\\\
{\rm Squark\ propagator:} \qquad & {\displaystyle \frac{1}{\hat{q}^2+m^2}}\\\\
{\rm Gluino\ propagator:} \qquad & {\displaystyle \frac{2}{i\,\qcircslash  + {2r\over a} \sum_\mu \sin^2(aq_\mu/2)}}\\
\end{array}
\ee

We have also listed the vertices which are required for carrying out the lattice computations. The extra vertices on the lattice are enumerated in Fig.~\ref{EXTRAlattice} and all vertices' expressions on the lattice are given in Eqs.~(\ref{qqGG})~-~(\ref{GGM}) in momentum space. In these expressions we have rescaled all momenta $k_i$ to the range  $[ - \pi,\pi ]$ and omitted overall powers of $a$.

\bea
\label{qqGG}
V_1(k_1,k_2,k_3) &=& i g (2\pi)^4 \delta(k_1-k_2+k_3) T^{\alpha_1}_{a_2a_3} \left(\gamma_{\mu_1} \cos\left(\frac{(k_2+k_3)_{\mu_1}}{2}\right) - i r  \sin\left(\frac{(k_2+k_3)_{\mu_1}}{2}\right) \right) 
{\phantom {1\over 2}}\\
V_2(k_1,k_2,k_3)   &=&   2 g (2\pi)^4 \delta(k_1-k_2+k_3) T^{\alpha_1}_{a_2 a_3} \sin\left(\frac{(k_2+k_3)_{\mu_1}}{2}\right)
{\phantom {1\over 2}}\\
V_3(k_1,k_2,k_3) &=& - 2 g (2\pi)^4 \delta(k_1-k_2+k_3) T^{\alpha_1}_{a_3 a_2}  \sin\left(\frac{(k_2+k_3)_{\mu_1}}{2}\right)
{\phantom {1\over 2}}\\
V_4(k_1,k_2,k_3)  &=&  \frac{1}{2} g (2\pi)^4 \delta(k_1-k_2+k_3) f^{\alpha_1 \alpha_2 \alpha_3} \left(\gamma_{\mu_1} \cos\left(\frac{(k_2+k_3)_{\mu_1}}{2}\right) -  i  r \sin\left(\frac{(k_2+k_3)_{\mu_1}}{2}\right) \right)
{\phantom {1\over 2}}\\
V_5(k_1,k_2,k_3)   &=& - 2\, i\, g \,(2\pi)^4 \delta(k_1-k_2+k_3)\, f^{\alpha_1 \alpha_2 \alpha_3} \cos\left(\frac{k_{3\,\mu_1}}{2}\right) \sin\left(\frac{k_{2\,\mu_1}}{2}\right)
{\phantom {1\over 2}}\\
V_6(k_1,k_2,k_3)  &=& - i \sqrt2 g (2\pi)^4 \delta(k_1-k_2+k_3) \frac{1 -  \gamma_5}{2} T^{\alpha_1}_{a_2 a_3}
{\phantom {1\over 2}}\\
V_7(k_1,k_2,k_3)  &=& - i \sqrt2 g (2\pi)^4 \delta(k_1-k_2-k_3) \frac{1 +  \gamma_5}{2} T^{\alpha_1}_{a_2 a_3}
{\phantom {1\over 2}}\\
V_8(k_1,k_2,k_3)   &=& i \sqrt2 g \delta(-k_1+k_2-k_3) \frac{1 +  \gamma_5}{2} T^{\alpha_1}_{a_3 a_2}
{\phantom {1\over 2}}\\
V_9(k_1,k_2,k_3) &=& i \sqrt2 g \delta(-k_1+k_2+k_3) \frac{1 -  \gamma_5}{2} T^{\alpha_1}_{a_3 a_2} 
{\phantom {1\over 2}}\\
V_{10}(k_1,k_2,k_3) &=& i g (2\pi)^4 \delta(k_1+k_2+k_3) \delta_{\mu_1\,\mu_2} f^{\alpha_1 \alpha_2 \alpha_3} \cos\left(\frac{k_{3\,\mu_1}}{2}\right) \sin\left(\frac{(k_1-k_2)_{\mu_3}}{2}\right)
{\phantom {1\over 2}}\\
V_{11}(k_1,k_2,k_3,k_4) &=& \frac{1}{2} g^2 (2\pi)^4 \delta(k_1+k_2-k_3-k_4) T^{\alpha}_{a_1\,a_3} T^{\alpha}_{a_2\,a_4}
{\phantom {1\over 2}}\\
V_{12}(k_1,k_2,k_3,k_4) &=& \frac{1}{2} g^2 (2\pi)^4 \delta(k_1+k_2-k_3-k_4) T^{\alpha}_{a_3\,a_1} T^{\alpha}_{a_4\,a_2}
{\phantom {1\over 2}}\\
V_{13}(k_1,k_2,k_3,k_4) &=& - g^2 (2\pi)^4 \delta(k_1-k_2+k_3-k_4) T^{\alpha}_{a_1\,a_2} T^{\alpha}_{a_4\,a_3}
{\phantom {1\over 2}}\\
V_{14}(k_1,k_2,k_3,k_4)   &=& g^2 (2\pi)^4 \delta(k_1+k_2-k_3+k_4) \delta_{\mu_1\,\mu_2} (T^{\alpha_1} T^{\alpha_2})_{a_3\,a_4} \cos\left(\frac{(k_3+k_4)_{\mu_1}}{2}\right)
{\phantom {1\over 2}}\\
V_{15}(k_1,k_2,k_3,k_4) &=& g^2 (2\pi)^4 \delta(k_1+k_2-k_3+k_4) \delta_{\mu_1\,\mu_2} (T^{\alpha_1} T^{\alpha_2})_{a_4\,a_3} \cos\left(\frac{(k_3+k_4)_{\mu_1}}{2}\right)
{\phantom {1\over 2}}\\
V_{16}(k_1,k_2,k_3,k_4) &=& - g^2 (2\pi)^4 \delta(k_1+k_2+k_3+k_4) {\rm{Tr}}(T^{\alpha_1}T^{\alpha_2}T^{\alpha_3}T^{\alpha_4})\times \nonumber\\\nonumber
&&  \hspace{-2.5cm}        \Big[ \delta_{\mu_1 \mu_2 \mu_3 \mu_4}\left(\frac{2}{3} -\frac{2}{3} \sum_{\rho} \cos\left(k_{1\,\rho}\right) + \frac{1}{2} \sum_{\rho} \cos\left(k_1+k_2\right)_{\rho} \right)\nonumber\\\nonumber
&& \hspace{-2.5cm}          + \delta_{\mu_1 \mu_2 \mu_3}\left (-\frac{4}{3}\sin\left(\frac{k_{4\,\mu_1}}{2}\right) \sin\left(\frac{k_{4\,\mu_4}}{2}\right) 
                  +2   \sin\left(\frac{k_{4\,\mu_1}}{2}\right) \sin\left(\frac{(2k_1+k_4)_{\mu_4}}{2}\right) 
                 +2   \sin\left(\frac{k_{4\,\mu_1}}{2}\right) \sin\left(\frac{(2k_3+k_4)_{\mu_4}}{2}\right) \right)\nonumber\\\nonumber
 &&    \hspace{-2.5cm}          + \delta_{\mu_1 \mu_2}\delta_{\mu_3 \mu_4}
                \left(\cos\left(\frac{(k_3+k_4)_{\mu_1}}{2}\right)\cos\left(\frac{(k_3+k_4)_{\mu_3}}{2}\right)
                 -2 \cos\left(\frac{(k_3-k_4)_{\mu_1}}{2}\right)\cos\left(\frac{(k_3+k_4)_{\mu_3}}{2}\right) \right)\nonumber\\
 && \hspace{-2.5cm}   
            + \delta_{\mu_1 \mu_3}\delta_{\mu_2\mu_4} \left(    \cos\left(\frac{(k_1-k_3)_{\mu_2}}{2}\right) \cos\left(\frac{(k_2-k_4)_{\mu_1}}{2}\right)  \right) \Big]
\\
V_{17}(k_1,k_2,k_3,k_4) &=& \frac{1}{2} g^2(2\pi)^4 \delta(k_1+k_2-k_3+k_4) \delta_{\mu_1\,\mu_2} (T^{\alpha_1} T^{\alpha_2})_{a_3 a_4} \times \nonumber\\\nonumber
&& \left (- i \gamma_{\mu_1} \sin\left(\frac{(k_3+k_4)_{\mu_1}}{2}\right) + r \cos\left(\frac{(k_3+k_4)_{\mu_1}}{2}\right)\right)
\\
V_{18}(k_1,k_2,k_3,k_4) &=& \frac{1}{4} g^2(2\pi)^4 \delta(k_1+k_2-k_3+k_4) \delta_{\mu_1\,\mu_2} f^{\alpha_1\alpha_3\alpha}f^{\alpha_2\alpha_4\alpha} \times \nonumber \\
&&  \left (-i \gamma_{\mu_1} \sin\left(\frac{(k_3+k_4)_{\mu_1}}{2}\right) + r  \cos\left(\frac{(k_3+k_4)_{\mu_1}}{2}\right)\right)
\\
V_{19}(k_1,k_2,k_3,k_4)  &=& -\frac{1}{3} g^2 \delta_{\mu_1\,\mu_2} (2\pi)^4 \delta(k_1+k_2-k_3+k_4) f^{\alpha_1\alpha_3\alpha} f^{\alpha_2\alpha_4\alpha} \sin\left(\frac{k_{3\,\mu_1}}{2}\right) \sin\left(\frac{k_{4\,\mu_1}}{2}\right)
\\
V_{20}(k_1,k_2)&=& \frac{1}{12} N_c \,g^2 \delta_{\mu_1\,\mu_2} (2\pi)^4 \delta(k_1+k_2)  {\rm{Tr}}(T^{\alpha_1}\,T^{\alpha_2})
\label{GGM}
\eea

Using the  2-pt Green's functions of each field, we can determine the corresponding renormalization factor. The first result presented here, Eq.~(\ref{GF2quarklatt}), is the lattice inverse quark propagator up to one loop; in this equation the quantity proportional to $1/a$ contributes to the additive renormalization of the quark mass (critical mass). In all lattice expressions the systematic errors (coming from an extrapolation to infinite lattice size of our numerical loop integrals) are smaller than the last digit we present.
\be
\langle \tilde \psi^B(q) \tilde{\bar{\psi}}^B(q') \rangle^{L}_{\rm{inv}} = (2\pi)^4 \delta(q-q') \Bigg\{ i \qslash \left[1  - \frac{g^2\,C_F}{16\,\pi^2} \left[ 12.8025 - 4.7920 \alpha + (2+\alpha)\log\left(a^2\,q^2\right)\right]\right] + \frac{g^2\,C_F}{16\,\pi^2}\frac{1}{a} 51.4347 \,r \Bigg\} +{\cal{O}}(a).
\label{GF2quarklatt}
\ee
From the one-loop correction to the quark propagator we obtain the multiplicative renormalization factor of the quark field. To this end, we use the renormalization condition of Eq.~(\ref{GG2condition}) which connects the bare inverse 2-pt Green's function on the lattice with the $\MSbar$ renormalized one. To avoid heavy notation we  have omitted coordinate/momentum arguments, as well as Dirac/flavor indices on $\langle\psi\,\bpsi \rangle$:
\be
\langle \psi^R\,{\bar{\psi}}^R \rangle_{\rm inv} = Z_\psi^{-1}\, \langle\psi^B\,{\bar{\psi}}^B \rangle^L_{\rm inv}\Big{|}_{a \to 0}\,.
\label{GG2condition}
\ee
The left-hand side equals the inverse Green's function in  Eq.~(\ref{GF2quark}) without the pole parts. From this equation we extract the renormalization for the quark field on the lattice:
\be
Z_\psi^{L,\MSbar} = 1 + \frac{g^2\,C_F}{16\,\pi^2} \left( -16.8025 + 3.7920 \alpha - (2+\alpha)\log\left(a^2\,\bar\mu^2\right) \right).
\ee
In Eq.~(\ref{GF2quarklatt}), just as in the corresponding equation in the continuum, terms with $\gamma_5$ cancel out at one-loop level. The same observation holds also for the gluino 2-pt function, see Eq.~(\ref{GF2gluinolatt}). This means that the $\psi_+$ and $\psi_- $ components of massless quarks do not mix under renormalization, unlike the case of the squark propagator, see Eq.~(\ref{GF2squarklatt}). The quark critical mass can be read from Eq.~(\ref{GF2quarklatt}):
\be
m^{quark}_{crit.} = \frac{g^2\,C_F}{16\,\pi^2} \frac{1}{a}  51.4347  \,r
\ee
This result is in agreement with Ref.~\cite{Skouropathis}.

The gluino inverse propagator is given at the one-loop order by:
\bea
\label{GF2gluinolatt}
\langle \tilde \lambda^{B}(q) \tilde{\bar{\lambda}}^{B}(q') \rangle^{L}_{\rm{inv}} &=& (2\pi)^4 \delta(q-q') \Big\{ \frac{i}{2} \, \qslash \Big[ 1 + \frac{g^2\,N_f}{16\,\pi^2}\left(1.9209 - \log\left(a^2\,q^2\right)\right)\\\nonumber
&& - \frac{g^2\, N_c}{16\,\pi^2} \left( 16.6444 - 4.7920\,\alpha  + \alpha \log\left(a^2\, q^2 \right)\right)\Big] + \frac{g^2}{16\,\pi^2} \frac{N_c}{2} \frac{1}{a} 51.4347 \,r\Big\}  +{\cal{O}}(a).
\eea

The renormalization factor of the gluino field is determined in the $\MSbar$ scheme by imposing the following condition:
\be
\langle\lambda^R\,{\bar{\lambda}}^{R} \rangle _{\rm inv} = Z_{\lambda}^{-1}\, \langle\lambda^B\,{\bar{\lambda}}^{B} \rangle^L_{\rm inv}\Big{|}_{a \to 0}
\label{GG2gluinocondition}
\ee
leading to:
\be
Z_{\lambda}^{L,\MSbar} =  1 - \frac{g^2\,}{16\,\pi^2} \left[N_c\left(16.6444 - 3.7920 \alpha +  \alpha \log\left(a^2\,\bar\mu^2\right)\right)+ N_f\left(0.07907 + \log\left(a^2\,\bar\mu^2\right)\right) \right],
\ee
and the critical mass for the gluino field is:
\be
m^{gluino}_{crit.} =  \frac{g^2\,N_c }{16\,\pi^2}\ \,\frac{1}{a}  51.4347 \,r 
\ee

In order to discuss the renormalization and mixing of squarks, it is convenient to write the bare squark fields in 2-component form:
\be
A^B \equiv \left( {\begin{array}{c} {A_+^B}^{\phantom{\dagger}} \\ {A_-^B}^{\dagger} \end{array} } \right).
\ee
The inverse squark propagator (without ${\cal O}(1/a^2)$ contributions), is given to one-loop order by:

\bea
\hspace{-1.25cm}\langle \tilde A^B(q) \tilde A^{B\,\dagger}(q') \rangle^{L}_{\rm{inv}} &=& (2\pi)^4 \delta(q-q') \Bigg\{ q^2 \begin{pmatrix} 1 & 0\\ 0 & 1 \end{pmatrix} - \frac{g^2\,C_F}{16\,\pi^2}q^2\Bigg[ \left[11.0173 - 3.7920 \alpha + (1+\alpha)\log(a^2\,q^2) \right] \begin{pmatrix} 1 & 0\\ 0 & 1 \end{pmatrix} \nonumber\\
&&\hspace*{6cm} +1.0087 \begin{pmatrix} 0 & 1\\ 1 & 0 \end{pmatrix}\Bigg]\Bigg\} +{\cal{O}}(a),
\label{GF2squarklatt}
\eea
and the critical mass for the squark fields (${\cal O}(1/a^2)$ contributions) is:

\be
m^{2\,squark}_{crit.}=- \frac{g^2\,C_F}{16\,\pi^2} \frac{1}{a^2} \begin{pmatrix}65.3930 & 75.4031 \\ 75.4031 & 65.3930
\end{pmatrix}.
\ee

We define the renormalization mixing matrix for the squark fields as follows:
\be
\label{condS}
\left( {\begin{array}{c} A^R_+ \\ A^{R\,\dagger}_- \end{array} } \right)= \left(Z_A^{1/2}\right)\left( {\begin{array}{c} A^B_+ \\ A^{B\,\dagger}_- \end{array} } \right).
\ee
Substituting Eq.~(\ref{condS}) into Eq.~(\ref{GF2squarklatt}), and requiring agreement with Eq.~(\ref{GF2squark}) in the $a\to 0$ limit, we find:
\be
\left(Z_A^{1/2}\right)^{L,\MSbar}= \openone - \,\frac{g^2\,C_F}{16\,\pi^2}\Bigg\{\Bigg[8.1753 - 1.8960\alpha+\frac{1}{2}(1+\alpha)\log\left(a^2\,\bar\mu^2\right)\Bigg] \begin{pmatrix} 1 & 0\\ 0 & 1 \end{pmatrix} - 0.1623 \begin{pmatrix} 0 & 1\\ 1 & 0 \end{pmatrix}\Bigg\}
.
\ee
The gluon inverse propagator is given to one loop by:
\bea
\langle  \tilde u_\mu^{B}(q) \tilde u_\nu^{B}(q') \rangle^{L}_{\rm{inv}} &=&(2\pi)^4 \delta(q+q') \Bigg\{ \frac{1}{\alpha}  q_{\mu} q_{\nu}\\\nonumber
&&+ \left(q^2 \delta_{\mu \nu} - q_{\mu} q_{\nu}\right)\Bigg[1- \frac{g^2}{16\,\pi^2}\Big[ -19.7392 \frac{1}{N_c} + N_f\left(-2.9622 + \log\left(a^2\,q^2\right)\right)\\\nonumber
&&+ N_c\left(20.1472 - 0.8863\,\alpha +  \frac{\alpha^2}{4} + \left(\frac{\alpha}{2} -\frac{3}{2}\right)  \log\left(a^2\, q^2 \right)\right)\Big]\Bigg]\Bigg\}+{\cal{O}}(a).
\label{GF2gluonlatt}
\eea
For $\langle  \tilde u_\mu^{B}(q) \tilde u_\nu^{B}(q') \rangle^{L}_{\rm{inv}}$, some diagrams contribute a quadratically divergent mass term ($1/a^2$ contribution). But when all Feynman diagrams are summed these divergences are found to cancel out. Another cancellation worthy of note regards non-covariant terms of type $(\delta_{\mu \nu} q^2_{\nu})$; after summing all contributions, these terms cancel out and one is left with a transverse expression for the gluon self energy, reflecting the gauge invariance of the theory. Since there is no critical mass or longitudinal part for the gluon self-energy, the renormalization of the gauge parameter $Z_\alpha$ receives no one-loop contribution. Our result for the gluon propagator without diagrams which involve squarks and gluinos, is consistent with Ref.~\cite{CMO}, where we calculate the same quantity in the non-supersymmetric case. 
By demanding the following:
\be
\langle u_\mu^R\,u_\nu^R \rangle _{\rm inv} = Z_{u}^{-1}\, \langle u_\mu^B\,u_\nu^B \rangle^L_{\rm inv},
\label{GG2gluoncondition}
\ee
we find:
\be
Z_{u}^{L,\MSbar} =  1 + \frac{g^2\,}{16\,\pi^2} \left[19.7392\frac{1}{N_c}- N_c\left(18.5638 - 1.3863  \alpha + \left( -\frac{3}{2}+\frac{\alpha}{2}\right) \log\left(a^2\,\bar\mu^2\right)\right)+ N_f\left(0.9622   - \log\left(a^2\,\bar\mu^2\right)\right)\right].
\ee
The ghost field renormalization, $Z_c$, which enters the evaluation of $Z_g$  can be extracted from the ghost propagator:
\be
\langle \tilde c^{B}(q) \tilde{\bar{c}}^{B}(q') \rangle^{L}_{\rm{inv}}  =  (2\pi)^4 \delta(q-q') q^2 \left[ 1-\frac{g^2\,N_c}{16\,\pi^2}\left(4.6086 - 1.2029 \alpha -\frac{1}{4}\left( 3 - \alpha \right)
\log\left(a^2\,q^2\right)\right)\right]+{\cal{O}}(a),
\label{GF2ghostlatt}
\ee
and $Z_c^{L,\MSbar}$ is:
\be
Z_c^{L,\MSbar} = 1 - \frac{g^2 N_c}{16\pi^2} \Bigl[3.6086 - 1.2029 \alpha -\frac{1}{4}\left( 3 - \alpha \right)
\log\left(a^2\,\bar{\mu}^2\right) \Bigr].
\ee

As in DR, we extract the coupling constant renormalization, $Z_g$, from the gluon-antighost-ghost Green's function $\langle  u_\mu^{B\,\alpha}(x) c^{B\,\beta}(x) \bar c^{B\,\gamma}(y) \rangle^{L}_{\rm{amp}}$\,. In the $\MSbar$ scheme, the renormalization condition, by analogy with Eq.~(\ref{GF3condition}), is:

\be
\lim_{a \to 0}\Big[\left(Z_{c}^{L,\MSbar}\right)^{-1} \left(Z_{u}^{L,\MSbar}\right)^{-1/2}\left(Z_{g}^{L,\MSbar}\right)^{-1}  \langle \tilde c^{B\,\alpha}(q) \tilde{\bar{c}}^{B\,\beta}(0) \tilde{u}_\mu^{B\,\gamma}(q')\rangle^{L}_{\rm{amp}} \Big] = \langle \tilde c^{B\,\alpha}(q) \tilde{\bar{c}}^{B\,\beta}(0) u_\mu^{B\,\gamma}(q')\rangle^{DR}_{\rm{amp}}\Big{|}_{1/\epsilon \to 0}
\label{GF3conditionlatt}
\ee
where the expression for $\langle \tilde c^{B\,\alpha}(q) \tilde{\bar{c}}^{B\,\beta}(0) u_\mu^{B\,\gamma}(q')\rangle^{DR}_{\rm{amp}}\Big{|}_{1/\epsilon \to 0}$ is the $\MSbar$-renormalized 3-pt Green's function which was calculated in the continuum, and the corresponding expression on the lattice is:

\be
\langle \tilde c^{B\,\alpha}(q) \tilde{\bar{c}}^{B\,\beta}(0) \tilde{u}_\mu^{B\,\gamma}(q')\rangle^{L}_{\rm{amp}}  = 
(2\pi)^4 \delta(q+q') f^{\alpha\,\beta\,\gamma} \left(i g  q_\mu\right) \Bigg[1 + \frac{g^2\,N_c}{16\,\pi^2}\left(
2.3960 \alpha - \frac{1}{2} \alpha \log\left(a^2\, q^2 \right)\right)
\Bigg]+{\cal{O}}(a).
\ee

Our result for $Z_g^{L,\MSbar}$ is:
\be
Z_g^{L,\MSbar} =  1 + \gtilde\,\Bigg[ -9.8696 \frac{1}{N_c} + N_c \left( 12.8904  - \frac{3}{2} \log\left(a^2\,\bar{\mu}^2\right)\right)-\,N_f\left( 0.4811 - \frac{1}{2} \log(a^2\,\bar{\mu}^2)\right)\Bigg].
\ee
From the calculation of $Z_{g}^{L,\MSbar}$ one can extract the Callan-Symanzik beta-function for SQCD. On the lattice the bare beta-function is defined as: 
\be
\beta_L(g^B)=-a\frac{dg^B}{da}|_{g^R,\,\bar{\mu}} 
\ee
In the asymptotic limit for SQCD, the expansion of the beta-function is done in powers of the bare coupling constant. The first term in this expansion is:
\be
\beta_L(g)=\frac{g^{3}\,}{16\,\pi^2} \left(-3N_c + N_f\right)+{\cal{O}}(g^5).
\ee

For $N_f < 3N_c$, the ${\cal{O}}(g^3)$ term is negative, in other words, the theory is asymptotically free. Our finding for the beta function agrees with what is obtained in the supersymmetric Yang-Mills theory~\cite{DRT}.

\section{Summary -- Conclusions}
\label{summary}

In this paper we have performed a pilot investigation of issues related to the formulation of a supersymmetric theory on the lattice. As a prototype model, we have  studied ${\cal N} = 1$ Supersymmetric QCD. This model bears all major characteristics of potential extensions of the Standard Model, including superpartners for gauge and matter fields; it is thus appropriate for a feasibility study on the lattice.

There are several well-known problems arising from the complete (or even partial) breaking of Supersymmetry in a regularized theory, including the necessity for fine tuning of the theory's bare Lagrangian, and a rich mixing pattern of composite operators at the quantum level. We address these problems via perturbative calculations at one loop. In order to provide the necessary ingredients for performing numerical studies of supersymmetric theories, we have calculated the self energies of all particles which appear in SQCD. We determined the renormalization factors for these fields; in addition, for the squark propagator we found the mixing coefficients among its different degrees of freedom. Furthermore, we have computed the gluon-antighost-ghost Green's function in order to renormalize the coupling constant. Our results are also relevant to the investigation of relationships between different Green's functions involved in SUSY Ward identities \cite{Taniguchi:2000, Vladikas:2002}.
   
There are several directions in which this work could be extended. A natural extension would be the computation of the Green's functions for composite operators made of quark, squark, gluon and gluino fields; studies of such operators in the continuum can be found in, e.g., Refs.~\cite{Konishi, Rattazzi, Silvestroni, Matthias}. A serious complication in the supersymmetric case regards the mixing of quark bilinear operators with other composite operators. A whole host of operators with equal or lower dimensionality, having the same quantum numbers and same transformation properties can mix at the quantum level; on the lattice, the number of operators which mix among themselves is considerably greater than in the continuum regularization. We are planning to study their renormalization and mixing perturbatively. The perturbative computation of all relevant Green's functions of these operators, will be followed by the construction of the mixing matrix, which may also involve non gauge invariant (but BRST invariant) operators or operators which vanish by the equations of motion. 

Finally, it would be important to extend our computations to further improved actions with reduced lattice artifacts and reduced symmetry breaking, e.g. the overlap fermion action, as a forerunner to numerical studies using these actions.


\begin{thebibliography}{99}

\bibitem{Feo:2003}
A.~Feo,
Nucl. Phys. Proc. Suppl. {\bf 119} (2003) 198, [hep-lat/0210015]. 


\bibitem{Giet&Poppitz}
J.~Giedt, E.~Poppitz, 
JHEP {\bf 9} (2004) 029, [hep-th/0407135]. 

\bibitem{Creutz:2001}
M.~Creutz, 
Rev. of Mod. Phys. {\bf 73} (2001) 119, [hep-lat/0007032]. 

\bibitem{Suzuki&Tani:2005}
H.~Suzuki, Y.~Taniguchi, 
JHEP {\bf 10} (2005) 082, [hep-lat/0507019]. 

\bibitem{Kaplan:2009}
S.~Catterall, D.B.~Kaplan, M.~\"Unsal,
Phys. Rep. {\bf 484} (2009) 71, [arXiv:0903.4881].

\bibitem{Catterall:2011} S.~Catterall, E.~Dzienkowski, J.~Giedt, A.~Joseph, R.~Wells, 
JHEP {\bf 4} (2011) 74, [arXiv:1102.1725]. 

\bibitem{Feo:2013}
A.~ Feo,
Phys. Rev. {\bf D88} (2013) 091501, [arXiv:1305.6473].

\bibitem{Catterall:2014}
S.~Catterall, D.~Schaich, P.H.~Damgaard, T.~DeGrand, J.~Giedt,
Phys. Rev. {\bf D90} (2014) 065013, [arXiv:1405.0644].

\bibitem{Curci&Venz}  G. Curci, G. Veneziano,
Nucl. Phys. {\bf B292} (1987) 555.

\bibitem{Giedt} J.~Giedt,
Int. J. Mod. Phys. {\bf A24} (2009) 4045, [arXiv:0903.2443]. 

\bibitem{josephREVIEW}
A.~Joseph, Int. J. Mod. Phys. {\bf A30} (2015) 1530054, [arXiv:1509.01440]. 

\bibitem{Grisaru:1983}
S.J.~Gates, M.T.~Grisaru, M.~Rocek, W.~Siegel,  {\em Superspace or One Thousand and One Lessons in Supersymmetry}, Front. Phys. {\bf 58} (1983) 1, [hep-th/0108200].

\bibitem{Wess&Bagger:1992}
J.~Wess, J.~Bagger, {\em Supersymmetry and Supergravity}, Princeton University Press (1992).

\bibitem{Weinberg:2000}
S.~Weinberg, {\em The Quantum Theory of Fields, Volume III Supersymmetry}, Cambridge University Press (2000).

\bibitem{MartinP}
S.~P.~Martin, {\em A Supersymmetry Primer}, Adv. Ser. Dir. High Energy Phys. {\bf 18} (1998) 1, [hep-ph/9709356].

\bibitem{Suzuki} H.~Suzuki,
Nucl. Phys. {\bf B861} (2012) 290, [arXiv:1202.2598]. 

\bibitem{Miller:1983} R.D.C.~Miller,
Phys. Lett. {\bf 129B} (1983) 72.

\bibitem{Jack&Jones}
I.~Jack, D.R.T.~Jones, High Energy Phys. {\bf 18} (1998) 149, [hep-ph/9707278].

\bibitem{Retey} K.~G.~Chetyrkin, A.~R\'etey, Nucl. Phys. {\bf B583} (2000) 3, [hep-ph/9910332];
[hep-ph/0007088].

\bibitem{Gracey} J.~A.~Gracey, 
Nucl. Phys. {\bf B662} (2003) 247, [hep-ph/0304113].

\bibitem{Hooft:1972}
G.~'t~Hooft, M.~Veltman, Nucl. Phys. {\bf B44} (1972) 189.

\bibitem{Larin}
S.A.~Larin, Phys. Lett. {\bf B303} (1993) 113, [hep-ph/9302240, containing an extra section].

\bibitem{Furman:2003}
M. Chanowitz, M.~Furman, I.~Hinchliffe, Nucl. Phys. {\bf B159} (1979) 225. 

\bibitem{Siegel:1979}
W.~Siegel, Phys. Lett. {\bf 84B} (1979) 193.

\bibitem{Patel}
A.~Patel, S.~R.~Sharpe, Nucl. Phys. {\bf B395} (1993) 701, [hep-lat/9210039].

\bibitem{buras} A.~J.~Buras, P.~H.~Weisz, 
Nucl. Phys. {\bf B333} (1990) 66.

\bibitem{HP} A.~Skouroupathis, H.~Panagopoulos, 
Phys. Rev. {\bf D76} (2007) 094514, [arXiv:0707.2906]. 

\bibitem{Gracey:2003}
J.A.~Gracey, Nucl. Phys. {\bf B662} (2003) 247, [hep-ph/0304113].

\bibitem{Feo} A.~Feo,	
Mod. Phys. Lett. {\bf A19} (2004) 2387, [hep-lat/0410012].


\bibitem{Skouropathis}
A.~Skouroupathis, M.~Constantinou, H.~Panagopoulos, Phys. Rev. {\bf D77} (2008) 014513, [hep-lat/0611005].

\bibitem{CMO} 
M.~Constantinou, M.~Costa, R.~Frezzotti, V.~Lubicz, G.~Martinelli, D.~Meloni, H.~Panagopoulos, S.~Simula, Phys. Rev. {\bf D92} (2015) 034505, [arXiv:1506.00361].

\bibitem{DRT} D.R.T.~Jones, Phys. Lett. {\bf 123B} (1983) 45.

\bibitem{Taniguchi:2000}
Y. Taniguchi,
Phys. Rev. {\bf D63} (2000) 014502, [hep-lat/9906026]

\bibitem{Vladikas:2002}   
F.~Farchioni, C.~Gebert, R.~Kirchner, I.~Montvay, A.~Feo, G.~M\"unster, T.~Galla, A.~Vladikas, 
Eur. Phys. Jour. {\bf C23} (2002) 719, [hep-lat/0111008]

\bibitem{Konishi}
K.~Konishi,	
Phys. Lett. {\bf 135B} (1984) 439.


\bibitem{Rattazzi}
G.F.~Giudice, R.~Rattazzi,
Phys. Rept. {\bf 322} (1999) 419, [hep-ph/9801271].


\bibitem{Silvestroni}
A.J.~Buras, P.~Gambino, M.~Gorbahn, S.~Jager, L.~Silvestrini,
Phys. Lett. {\bf B500} (2001) 161, [hep-ph/0007085].


\bibitem{Matthias}
C.~Greub, T.~Hurth, V.~Pilipp, C.~Sch\"upbach, M.~Steinhauser,  	
Nucl. Phys. {\bf B853} (2011) 276, [arXiv:1105.1330].



\end{thebibliography}
\end{document}